\documentclass{optica-article}

\journal{opticajournal} 

\articletype{Research Article}

\usepackage{xr}
\usepackage{soul}
\usepackage{amsmath}
\usepackage{graphicx}
\usepackage{epstopdf}
\usepackage{subfig}
\usepackage{multirow}
\usepackage{url}
\usepackage{float}
\usepackage{bm}
\usepackage{ifthen}
\usepackage{braket}

\usepackage{siunitx} 

\DeclareMathOperator{\Tr}{Tr}

\usepackage[absolute]{textpos}
\usepackage{xcolor}
\usepackage{placeins}

\begin{document}

\title{Continuous Automatic Polarization Channel Stabilization from Heterodyne Detection of Coexisting Dim Reference Signals}

\author{Joseph C. Chapman,\authormark{1,*} Muneer Alshowkan,\authormark{1}, Kazi Reaz,\authormark{2} Tian Li,\authormark{2} and Mariam Kiran\authormark{1}}

\address{\authormark{1} Quantum Information Science Section, Oak Ridge National Laboratory, Oak Ridge, TN 37831, USA}
\address{\authormark{2} Department of Physics and Chemistry and UTC Quantum Center, University of Tennessee at Chattanooga, Chattanooga, TN 37403, USA}

\email{\authormark{*}chapmanjc@ornl.gov} 


\begin{abstract*} 
Quantum networking continues to encode information in polarization states due to ease and precision. The variable environmental polarization transformations induced by deployed fiber need correction for deployed quantum networking. Here we present a new method for automatic polarization compensation (APC) and demonstrate its performance on a metropolitan quantum network. Designing an APC involves many design decisions indicated by the diversity of previous solutions in the literature. Our design leverages heterodyne detection of wavelength-multiplexed dim classical references for continuous high-bandwidth polarization measurements used by newly developed multi-axis (non-)linear control algorithm(s) for complete polarization channel stabilization with no downtime. This enables continuous relatively high-bandwidth correction without significant added noise from classical reference signals. We demonstrate the performance of our APC using a variety of classical and quantum characterizations. Finally, we use C-band and L-band APC versions to demonstrate continuous high-fidelity entanglement distribution on a metropolitan quantum network with average relative fidelity of $0.94\pm0.03$ for over 30 hrs.
\end{abstract*}


\section{Introduction}
Deployed fiber transforms a transmitted signal’s polarization differently depending on its local environment via local stress-induced birefringence. In communications, this was first an issue for the advancement of classical coherent communications~\cite{ulrich1979polarization,Okoshi_1985} and later for quantum communications~\cite{RevModPhys.74.145}. Stabilizing the polarization state always involves having some known reference polarization state transmitted through the channel and (at least partially) measuring the polarization at the output of the channel and using that measurement result to compensate for the polarization changes. 

There are different kinds of stability required for different applications so not all polarization compensation is directly comparable. Some applications only need the transmitted power through a polarizer to be stable, whereas others need the state of polarization (SOP) of a certain signal to be stable. These two operating modes are common in classical communications and sensing where there is a defined polarization state desired which can be partially measured classically as part of compensation and the rest transmitted for the desired application. Finally, others need the complete polarization transformation (i.e., polarization channel) of the fiber for any input polarization to be stable. This is the case for quantum communications where polarization entanglement is in all polarization bases. The difference between stabilizing a certain SOP and stabilizing the complete polarization transformation is subtle and comes down to rotation sensitivity. Specifically, when using one reference polarization state to stabilize a certain SOP, the measurement is insensitive to polarization rotations on the Poincare sphere about that measured SOP. Thus, the distinguishing factor between stabilizing a single SOP and stabilizing the complete polarization transformation comes down to one or multiple reference polarizations.

With that in mind, the main design choices are (1) What is the compensation goal? (2) What is the light source, wavelength, and power of the reference signal(s)? and (3) What is used to detect the reference after transmission through the channel? (4) How is the polarization adjustment determined using the detector output? (5) How is the polarization adjusted?

For classical communications, a variety of solutions have been proposed to control the polarization of signals transmitted through fiber. For choice (1), as previously mentioned, classical communications focuses on stabilizing the output of a polarizer or a single SOP. For choice (2), lasers are ubiquitous. But otherwise, the solutions vary widely in their design choices. For choice (3), standard polarization optics (e.g., polarizer or polarimeter) and photodiodes~\cite{ulrich1979polarization,Giese_Schatzel,Noe_1986,Martinelli2006,Pikaar_van1989,Chiba1999,Zhao2009,Xiao2010}, whereas an alternative to photodiodes is coherent detection~\cite{Imai_1985} and recently integrated optics~\cite{Liu2022} instead of free-space- or fiber-based. For choice (4), the algorithms are often based on a type of global search~\cite{Noe_1986,Pikaar_van1989,Zhao2009,Xiao2010,Liu2022} but linear control, (e.g., proportional-integral-derivative) has also been demonstrated~\cite{ulrich1979polarization,Giese_Schatzel,Martinelli2006}. For choice (5), the manipulation is commonly done with fiber squeezers~\cite{ulrich1979polarization,Giese_Schatzel,Noe_1986,Pikaar_van1989,Zhao2009,Xiao2010} but also liquid-crystal variable retarders~\cite{Martinelli2006}, rotating wave-plates~\cite{Imai_1985,Chiba1999}, variable magneto-optical retarders~\cite{Martinelli2006}, and more recently integrated-optic interferometer polarization converter~\cite{Liu2022}.

There are several demonstrated methods to implement polarization stabilization of a fiber-based connection for quantum communications which have some similarities and some differences to the classical solutions mentioned above. For design choice (1), when homodyne detection is involved the goal can be to maximize transmission through a polarizer, whereas if the quantum signal is used as a polarization reference then the goal can be to stabilize the SOP but otherwise polarization channel stabilization is needed most likely. Similar to classical methods, it is common to make the goal centered around optimizing the error rate~\cite{IDref11,IDref12,IDref13,Peranic2023} or detected power through a polarizer~\cite{IDref6,IDref10,Wang2020}, whereas some others stabilize the quantum polarization state~\cite{IDref8,Chen2007}. Some use the quantum signal itself~\cite{IDref10,IDref11,IDref12,IDref13,Peranic2023} but sometimes other references are used~\cite{IDref6,IDref7,IDref8,Chen2007}. Notably, Ref.~\cite{Xavier2008,IDref9} use multiple polarization references to stabilize the complete polarization channel.

For quantum communications, choices (2) and (3) are intimately related so we will discuss them in tandem. The best performance (simplicity, accuracy, and bandwidth) is likely to be obtained by using a mW-power laser with a low-noise photodiode for the reference signal and detection, respectively, due to shot-noise limitations at low power. However, a mW-power (0~dBm) laser near the dim quantum signal (e.g., -120~dBm) will produce a significant amount of spontaneous Raman scattering at the wavelength of the quantum signal for any significant fiber propagation distance~\cite{IDref1,IDref2}. Previous results have shown acceptable coexistence with the classical signal and the discrete-variable (DV) quantum signal separated by several hundred nanometers~\cite{IDref3,IDref4,IDref5} but due to polarization-mode dispersion (PMD) having the polarization reference and the quantum signal so far separated in wavelength is untenable. In this situation, there are two main options to explore for nearby wavelength references: (a) time-multiplexing and (b) low-power reference and detection.

Time-multiplexing could allow for having the mW-power reference signal and the quantum signal on at different times (or ignoring the quantum signal when the reference signal is on)~\cite{IDref6,IDref7,IDref8,IDref9}. Time-multiplexing can beneficially allow the use of the mW-power reference signal and low-noise photodiode. A downside of time-multiplexing is that the correction bandwidth is limited by the repetition rate of the time-multiplexing; but if that repetition is faster than the environmental polarization transformations then that can be acceptable. Another downside of time-multiplexing the reference and quantum signal is that if the quantum signal is not also pulsed but is continuous then the duty cycle of operation (up-time) is reduced by the time-multiplexing leading to a reduced rate of quantum communication (e.g., entanglement distribution rates) and additional time synchronization requirements---which could become quite complex in a operational multi-node network.

For a low-power reference, there are several options which can be used: a significantly attenuated laser, a part of the quantum signal which is picked-off, or a polarization reference-operation mode for the quantum light source. Using an attenuated laser has the benefit of being adjustable to increase the power to maximize use of the dynamic range of any single-photon detectors used while being simple to operate and not affecting the quantum signal. However, combining a reference signal with the quantum signal and separating them will add insertion loss on the quantum signal reducing the entanglement distribution rate. When sending polarized quantum signals, it is possible to pick off some of the light in some way to measure the polarization using the quantum signal itself, e.g., Ref.~\cite{IDref10}. Using the quantum signal directly is also common in quantum cryptography demonstrations where error rates are regularly measured and are readily usable for polarization correction~\cite{IDref11,IDref12,IDref13}. When using an entangled photon source, the output is essentially unpolarized or randomly polarized one might say; it is a polarization entangled state without a definite polarization on each side. Thus, a direct measurement of an entanglement source cannot be used to stabilize the polarization. Additionally, some quantum light sources may be operated sometimes emitting entangled photons and other times emitting polarized photons but action of switching operation modes may lose the phase relationship of the entangled state.

When using a conventional photodiode with a mW power laser, photodiodes with bandwidths up to 50~GHz can be purchased leading to no constraint on the control system bandwidth. But if single-photon detectors are used, the input signal flux and the detectors dynamic range need to be considered together. Drawing on our comprehensive industry knowledge, we can conservatively assume a dynamic range of 40~dB (100~cps to 1~Mcps) for the single-photon detectors. The input signal flux is going to be signal and channel dependent. For single-photon detectors, the best-case scenario is when an attenuated laser is used which can be adjusted so the power is always near the max dynamic-range of the single-photon detector. 

The fourth design choice to discuss is the choice of polarization correction algorithm. In many of these cases, some optimization algorithm is used to reach the desired operating point. Global optimization solvers (genetic algorithms~\cite{IDref14}, particle swarm~\cite{IDref15}, simulated annealing~\cite{IDref16}, etc.) often require many samples and iterations to converge. Time for this optimization may or may not be acceptable, depending on the sampling and computation rate compared to the polarization rate of change. On the other hand, others have put forward analytical correction algorithms~\cite{IDref7,Ramos2020}. Linear control using proportional-integral-derivative error is ubiquitous across many applications of control theory for its simplicity and good performance. But it does not maintain continuous stability using setpoints near peaks or valleys or systems which dynamically change feedback sign (i.e., as in limited range trigonometric functions describing polarization). It has been applied to early studies of classical polarization SOP control~\cite{ulrich1979polarization,Giese_Schatzel} but not to quantum communication or full polarization channel control to our knowledge. Other algorithms to consider are the guess-and-check algorithm~\cite{IDref6} and sample-and-fit algorithm~\cite{IDref18}.

For design choice (5), there are six main fiber-based hardware options that can all rotate the polarization sufficiently well but they differ greatly in their speed and insertion loss. Table~\ref{tab:polrotmethds} compares these methods. The options in Table~\ref{tab:polrotmethds} span 11 orders of magnitude for the maximum bandwidth but there are tradeoffs with the monetary cost and insertion loss. The fiber-coupled motorized wave-plates, liquid-crystal variable retarders, and free-space electro-optic modulator (EOM) each have an insertion loss < 0.5 dB per element but there is a 0.5-1 dB fiber-coupling insertion loss, so the loss quoted is an aggregate loss assuming multiple elements in series are needed. The paddle rotators and fiber squeezers each manipulate a fiber directly to achieve the polarization rotation leading to very low insertion loss and each have multiple actuators so only one module is needed for polarization compensation per fiber. Though the paddle rotator insertion loss can be high in cases when a broadband device has loops small enough to cause excess loss for longer wavelengths. The waveguide EOM has the highest bandwidth but also the highest loss (2-4 dB is per element).

\begin{table}
\centering
\caption{Comparison of available fiber-based polarization rotation methods. \$ is $<$ \$1k. \$\$ is $<$ \$10k. \$\$\$ is $<$ \$100k. EOM = electro-optic modulator. LCVR = liquid-crystal variable retarders.}
\label{tab:polrotmethds}
\begin{tabular}{m{0.25\textwidth}m{0.2\textwidth}m{0.15\textwidth}m{0.15\textwidth}m{0.05\textwidth}}
\hline
Description & Bandwidth & Retardance & Insertion Loss & Cost\\
\hline
Fiber-coupled motorized wave plates & DC to 0.1 Hz & $\leq2\pi$ & 1-2 dB & \$\$\\
Manual or motorized fiber-loop paddle rotator &	DC to 10 Hz & $\leq2\pi$ &	0 - 15 dB &	\$ \\
Fiber-coupled LCVR &	DC to 200 Hz  & $\leq12\pi$ & 1-2 dB &	\$\$ \\
Piezo fiber squeezers &	DC to 20 kHz & $\leq5\pi$ &	 0.01 dB &	\$\$ \\
Fiber-coupled free-space EOM &	DC to  1 MHz	& $\leq2\pi$ & 1-2 dB	& \$\$\$ \\
Fiber-coupled waveguide EOM &	DC to 10 GHz & $\leq2\pi$ &2-4 dB	& \$\$\\
\hline
\end{tabular}
\end{table}

To make a good decision among all the design choices discussed above, it is important to know the bandwidth of polarization transformations imposed on the quantum signal by the fiber for a wide range of environmental conditions. Then the loop bandwidth of the polarization compensation would ideally be >10-100x the fiber transformation bandwidth so the corrections can be calculated and applied fast enough to maintain stable link performance throughout changing environmental conditions. Picking a compensation system (reference source, detection, compensation algorithm, and polarization actuation) with sufficient bandwidth is the first priority, then optimizing for low insertion loss and cost among the remaining choices. 

In this work, our aim is to enable fast complete polarization channel control with 100\% up-time compatible with quantum and classical communications by using dim (-50 dBm) {wavelength}-multiplexed reference signals, which are ultimately detected using a heterodyne spectrometer. Heterodyne spectrometry enables the use of high-bandwidth photodiodes in conjunction with near single-photon (per mode) sensitivity for high SNR and distinguishable references of nearly identical wavelengths. Moreover, we develop a compensation and measurement system in conjunction with a newly developed multi-axis control system to stabilize the complete polarization channel experienced by any input polarization. Moreover, the reference signals have sufficiently low power to have negligible added noise for classical and continuous-variable quantum communications with minimal filtering (i.e., single wavelength demultiplexer). For single-photon quantum communications, with moderate filtering, the Raman noise generated is negligible for modest coincidence windows.

{Compared to approaches which use single-photon detectors, this method enables significantly higher bandwidth detection and control by using heterodyne detection based on high-speed photodiodes. By wavelength-multiplexing dim reference signals with the input signal, there is no down-time compared to approaches which time-multiplex bright reference signals. Moreover, the control system we develop provides deterministic linear control (with a non-linear control variation for extended range) which continuously applies correction so to maintain the polarization channel which is advantageous to general search-based algorithms which periodically will search for optimal polarization state (or channel), sometimes temporarily worsening the polarization state in the process, and often without any control in between periodic checks. These are some of the benefits of our approach, for which more are described in Sec.~\ref{sec:discuss}. Moreover, a}lthough our work has similarities to Ref.~\cite{ulrich1979polarization,Imai_1985,Martinelli2006,Xavier2008}, it differs substantially in the stability goal and/or various design choices leading to superior overall performance and versatility.

Although, our implementation uses wavelength-division multiplexing, this compensation and detection scheme could also be used with temporal multiplexing of the references having the same wavelength as the quantum signal {if an application requires}. When wavelength-division multiplexing is used, a weakness of this scheme is its susceptibility to polarization-mode dispersion, which we address below. We previously studied heterodyne spectrometers~\cite{IDref19} and their use here highlights the bright narrow-bandwidth (reference) signals we found would be advantageous to use heterodyne spectrometers to detect for quantum networking. 


\section{Methods}
\label{sec:methods}
\subsection{Automatic Polarization Compensation}
\label{sec:methodsAPC}
Here we start by presenting an algorithm {(and an algorithm variation for extended range)} specifically suited for polarization stabilization using variable wave-plates (VWP), e.g., fiber squeezers or liquid-crystal variable retarders. The {feedback-algorithm-stage for both the main algorithm and its variation} includes a 3-dimensional 3-independent-loop proportional-integral-derivative (PID) control algorithm (implemented on several Liquid Instruments Moku:Go) use 3 total measurements on 2 total reference polarizations. {The presented algorithm is flexible to apply to all three major compensation goals, i.e., stabilize projection through polarizer, stabilize a single SOP using measurements on that SOP, and stabilize the polarization channel for any SOP input. But in this work, we focus on stabilizing the polarization channel.}

\begin{figure}
\centerline{\includegraphics[width=0.5\textwidth]{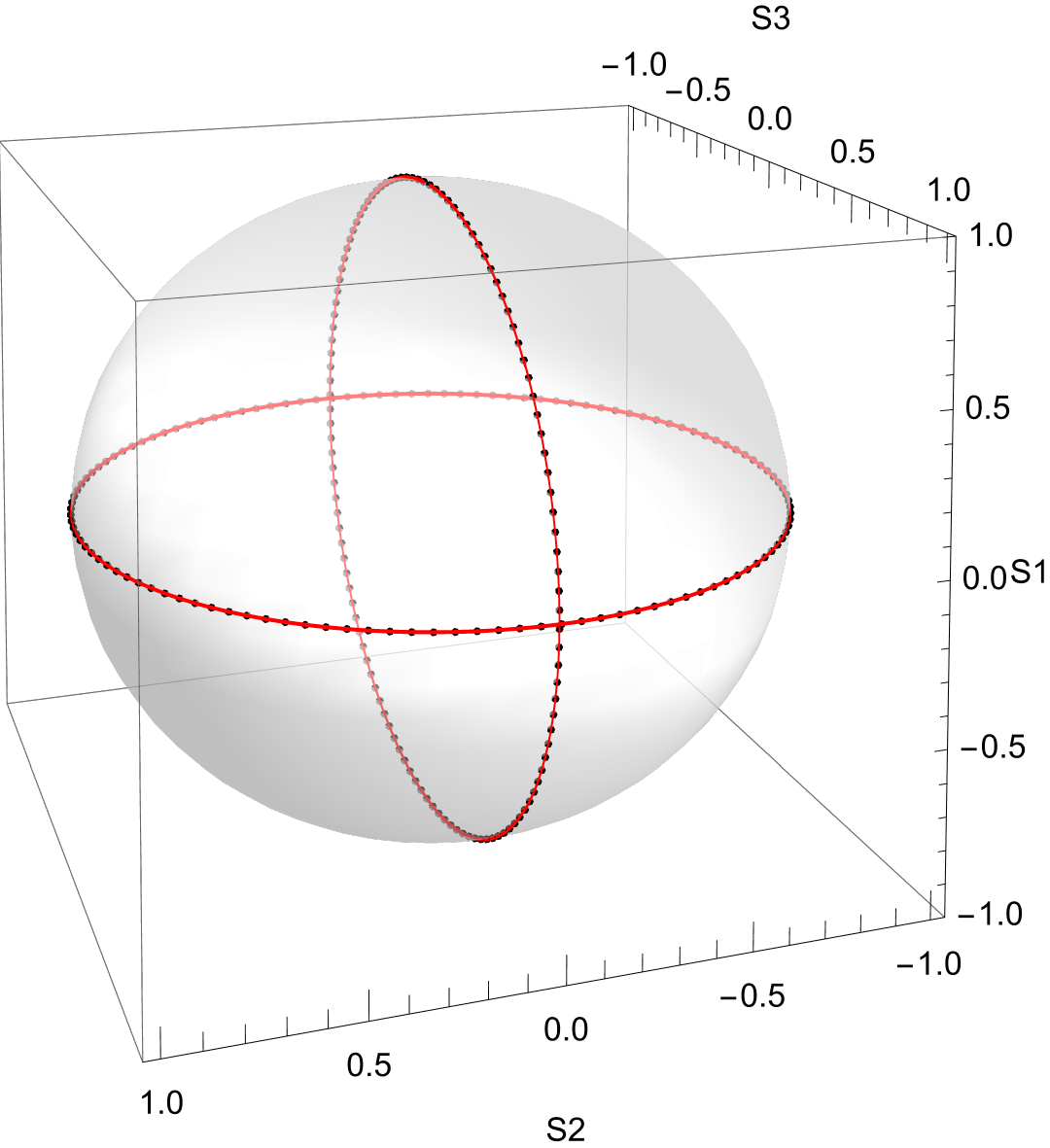}}
\caption{Poincare Sphere plotted in Stokes Vector Cartesian basis. Variable wave-plates in H/V and D/A basis each make great circles which are orthogonal with respect to each other in the vicinity of R and L.}
\label{fig:poincaresph}
\end{figure}

The algorithms are motivated by the action of a VWP on light’s polarization. Changing the phase of a VWP rotates the polarization of light along a great circle with the rotation axis being the polarizations that are eigenstates of the slow/fast axes of the VWP. For example, changing the phase of a VWP with the slow/fast axes in the horizontal/vertical (H/V) basis will take diagonal (D) input polarization to right circular (R), then anti-diagonal (A), then left circular (L), and finally back to D. Accordingly, a VWP in the H/V basis and another one in the D/A basis will both transform R input polarization but the great circles are practically orthogonal to one another at R in the Poincare sphere representation of polarization (Fig.~\ref{fig:poincaresph}). Since the H/V-basis VWP moves R along the D/A basis (as measured by Stokes $S_2$ = D - A), measuring R in the D/A basis provides information pertinent to the H/V-basis VWP. Similarly, since the D/A-basis VWP moves R to the H/V basis (as measured by Stokes $S_1$ = H - V), measuring R in the H/V basis provides information pertinent to the D/A basis VWP and these two measurements are orthogonal on the Poincare sphere. Given an R-reference polarization, a D/A-basis measurement after propagating through some channel (and any compensating elements, e.g., VWP) can be used with a linear control algorithm (e.g. PID) to adjust the H/V-basis VWP to constrain polarization after the compensating elements. Simultaneously, a H/V-basis measurement on the R-reference polarization can be used with a linear control algorithm to adjust the D/A-basis VWP to further constrain the polarization after the compensating elements. {In a generalized form, to control polarization using a VWP with rotational axis about basis 1, requires a measurement in basis 2 on a reference signal with polarization in basis 3, where basis 1, 2, and 3 can be any set of 3 mutually unbiased bases picked without duplication, i.e., basis 1, 2, and 3 are all different for a given control loop.}

\begin{figure}
\centerline{\includegraphics[width=1\textwidth]{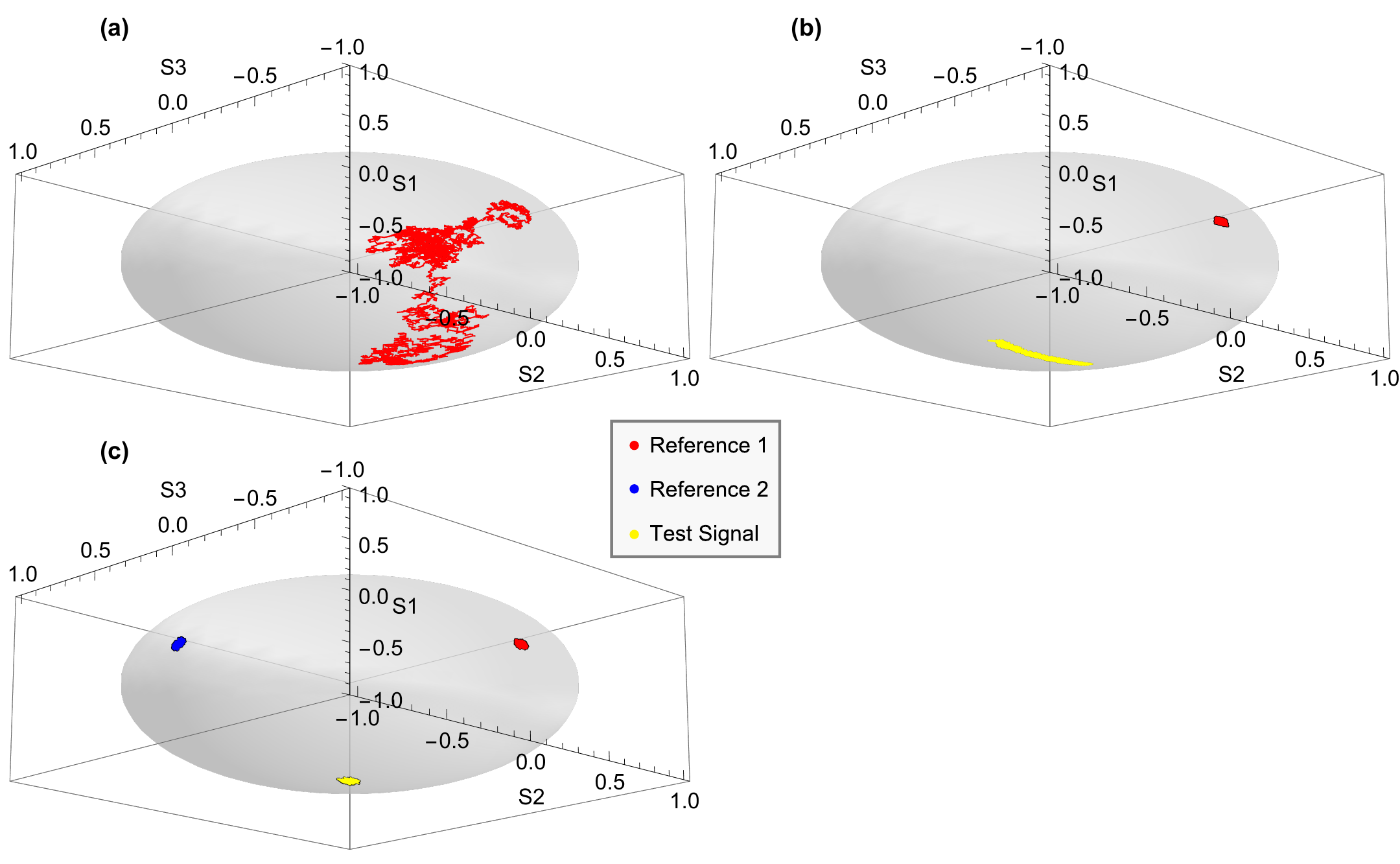}}
\caption{Algorithm simulation comparison using the Poincare sphere representation. (a) Reference 1 (starting at L) uncontrolled polarization drift simulated using a multi-coordinate random walk of $\pi/360$ (0.5$^{\circ}$) step sizes controlling the phase and angle of several variable wave-plates. (b) Reference 1 and V test signal controlled by 2-channel PID using H/V- and D/A-basis measurements on reference 1. (c) Reference 1, Reference 2 (starting at A), and V test signal controlled by 3-channel PID using H/V- and D/A-basis measurements on reference 1 and H/V-basis measurement on reference 2. The simulations were done using Mathematica~\cite{Mathematica}.}
\label{fig:algorithmcomp}
\end{figure}

If the only interest was to keep this reference at R (or L since there is ambiguity in these measurements), these measurements would be sufficient for polarization control of the reference signal [Fig.~\ref{fig:algorithmcomp}(b)] compared to the uncontrolled in [Fig.~\ref{fig:algorithmcomp}(a)]. But just these two measurements on one reference does not achieve complete polarization channel stability for any input polarization as needed in this application of quantum communication; there is ambiguity for rotations along the great circle with the R rotational axis [Fig.~\ref{fig:algorithmcomp}(b)]. To achieve complete channel control then another reference is needed that is not orthogonal to R, e.g., the simulations use A polarization as a second reference. Measuring this reference in the H/V basis (or D/A basis for an H or V basis second reference) and using that measurement to control an R/L basis VWP provides the needed additional control [Fig.~\ref{fig:algorithmcomp}(c)]. This control is also orthogonal to the others on the Poincare sphere in the vicinity of the setpoint so the three control loops can operate mostly independently. Thus, a complete set of compensating elements could include a VWP in the H/V basis followed by a VWP in the D/A basis, followed by a VWP in the R/L basis [e.g., realized by a quarter-wave plate (QWP) in the H/V basis then a VWP in the D/A basis then another QWP in the H/V basis but rotated 90 degrees with respect to the other QWP)]. In practice, several ways this can be realized are with free-space wave-plates and liquid-crystal variable retarders or a 4-element fiber squeezer device (the last QWP can be somewhere before the measurement) which we have set up and tested both variations. We have tested polarization stabilization with both types of compensating elements; we found the fiber squeezer performed markedly better in our initial testing due to better phase-shift linearity with applied voltage so in the following we will only discuss results using fiber-squeezer-based correction.

\begin{figure}
\centerline{\includegraphics[width=1\textwidth]{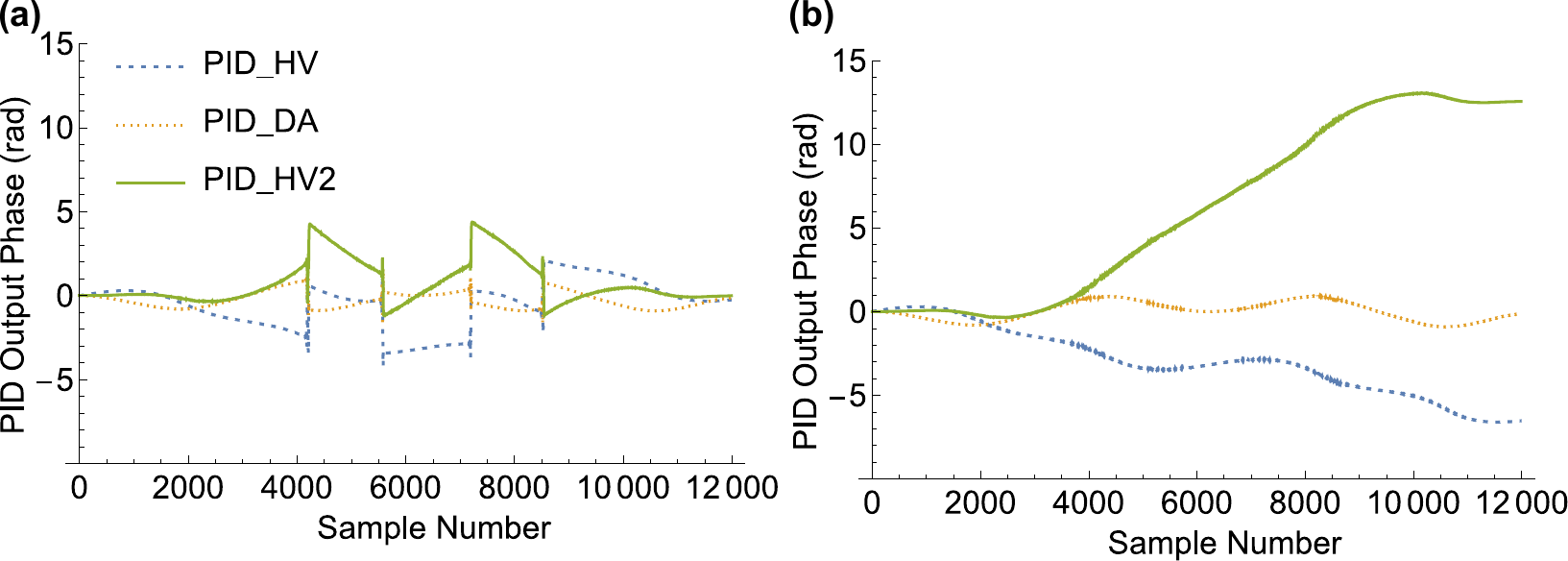}}
\caption{Algorithm comparison of PID outputs with biased random walk using $\pi/5000$ (0.036$^{\circ}$) steps. (a) Outputs of 3-channel PID using H/V- and D/A-basis measurements on reference 1 and H/V-basis measurement on reference 2. (b) Outputs of 3-channel PID with slope sign in error using H/V- and D/A-basis measurements on reference 1 and H/V-basis measurement on reference 2. The simulations were done using Mathematica~\cite{Mathematica}.}
\label{fig:pidoutalgthmcomp}
\end{figure}

\begin{figure}
\centerline{\includegraphics[width=1\textwidth]{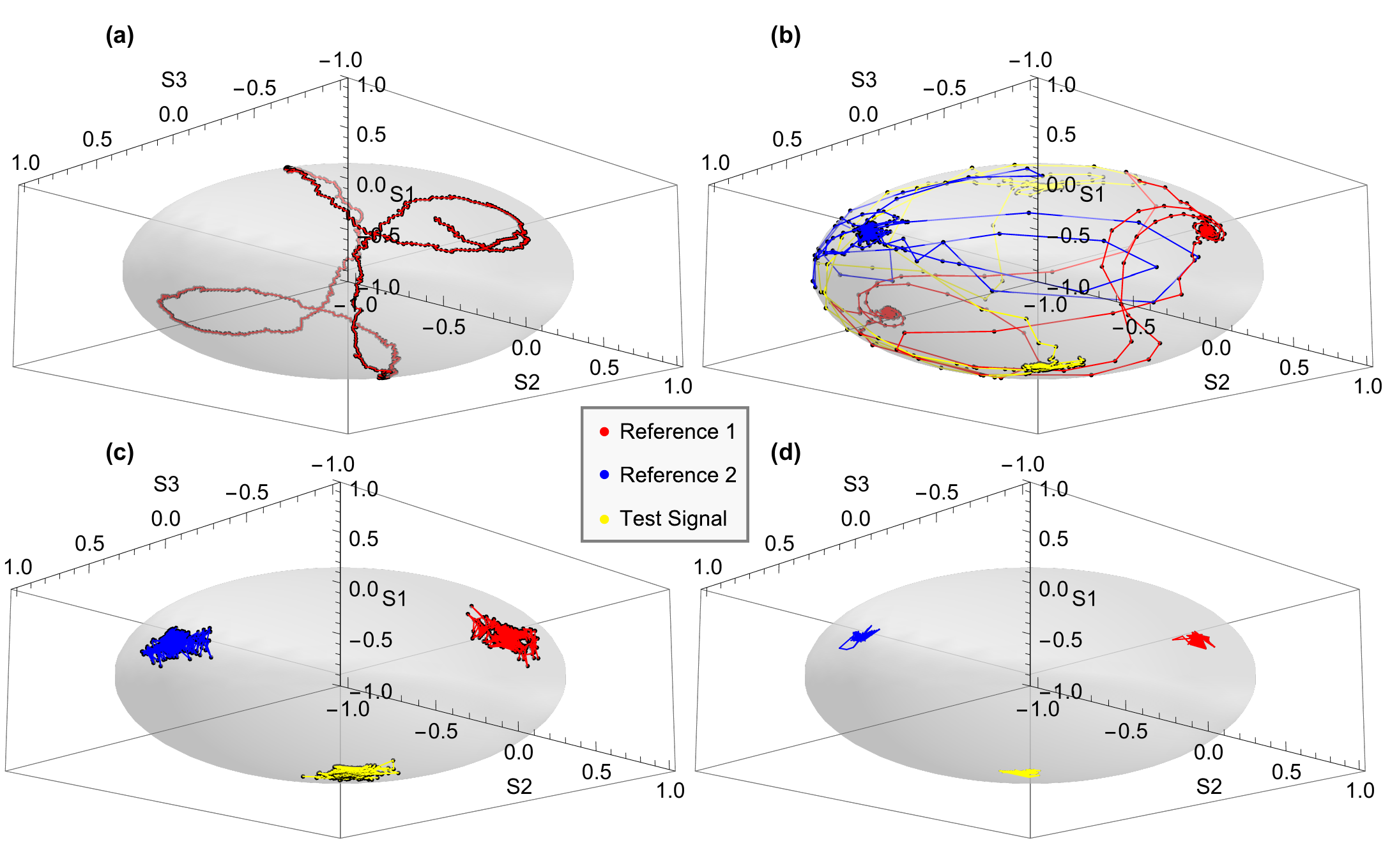}}
\caption{Slope-sign algorithm comparison using Poincare sphere representation. (a) Reference 1 (starting at L) uncontrolled polarization drift simulated using a multi-coordinate biased random walk of $\pi/360$ (0.5$^{\circ}$) step sizes controlling the phase and angle of several variable wave-plates. (b) Reference 1, Reference 2 (starting at A), and V test signal controlled by 3-channel PID using H/V- and D/A-basis measurements on reference 1 and H/V-basis measurement on reference 2. (c) Reference 1, Reference 2 (starting at A), and V test signal controlled by 3-channel PID with slope sign in error using H/V- and D/A-basis measurements on reference 1 and H/V-basis measurement on reference 2. (d) Reference 1, Reference 2 (starting at A), and V test signal controlled by 3-channel PID with slope sign in error using H/V- and D/A-basis measurements on reference 1 and H/V-basis measurement on reference 2 during random walk with smaller $\pi/5000$ (0.036$^{\circ}$) step sizes. The simulations were done using Mathematica~\cite{Mathematica}.}
\label{fig:slpsgnalgthmcomp}
\end{figure}

In simulation and experimental testing, we have found there is a range limitation of about $\pi$ for this control method due to a sign change of the slope of the Stokes measurement as a function of the variable waveplate applied retardance as the polarization would drift around the sphere which can be seen by the nearly discontinuous jumps in Fig.~\ref{fig:pidoutalgthmcomp}(a) and the bi-stability in Fig.~\ref{fig:slpsgnalgthmcomp}(b); This is discussed in more detail in Ref.~\cite{Martinelli2006}. This is not an issue for a small drift range as evidenced by Fig.~\ref{fig:algorithmcomp}(c), but if wrapping around the Poincare sphere is expected then the range can be extended with a variation on the base algorithm. To maintain control despite the slope-sign change, the input error of the PID algorithm can be modified to include this slope sign. Inspired by Ref.~\cite{IDref17}, the input error $e_n$ on iteration $n$ is modified from $e_n = (M_n-T)$, where $M_n$ is the current measurement and $T$ is the target setpoint, to
\begin{equation*}
e_n =
- (M_n-T) \begin{cases}
\text{Sgn}\Big(\frac{(M_n-T) - (M_{n-1}-T)}{C_{n-1}-C_{n-2}}\Big)& \text{if } C_{n-1}>C_{n-2},\\
\text{Sgn}\Big(\frac{(M_{n-1}-T) - (M_{n}-T)}{C_{n-2}-C_{n-1}}\Big) & \text{if } C_{n-1}<C_{n-2},
\end{cases}
\end{equation*}
where $C_{n}$ is the PID output correction applied after PID iteration $n$ using $M_n$ and $\text{Sgn}()$ is the sign function. This modification provides the linear control with information about the slope of the Stokes measurements as a function of the variable waveplate applied retardance using a finite difference to approximate the derivative. This algorithm variation thus becomes a non-linear controller but retains some of the benefits of linear control found in PID, namely, it retains most of the reliable continuous deterministic control. Using this improved algorithm, control for polarization drift anywhere on the sphere is achieved [Fig.~\ref{fig:slpsgnalgthmcomp}(c)]. This algorithm is necessarily noisier than regular PID because it can confuse signal noise for slope changes but it can provide complete polarization channel control up to the retardance limit of the variable wave-plates used. This is mostly apparent near the slope changes as evidenced by the somewhat noisier areas in Fig.~\ref{fig:slpsgnalgthmcomp}(b) which correspond to sign-change-induced jumps in Fig.~\ref{fig:slpsgnalgthmcomp}(a). In these simulations, the error can be reduced with smaller random walk step sizes [Fig.~\ref{fig:slpsgnalgthmcomp}(d)] showing that if the Stokes measurement SNR and polarization actuator precision is good enough then precise control can be achieved with this algorithm.

\begin{figure}
\centerline{\includegraphics[width=1\textwidth]{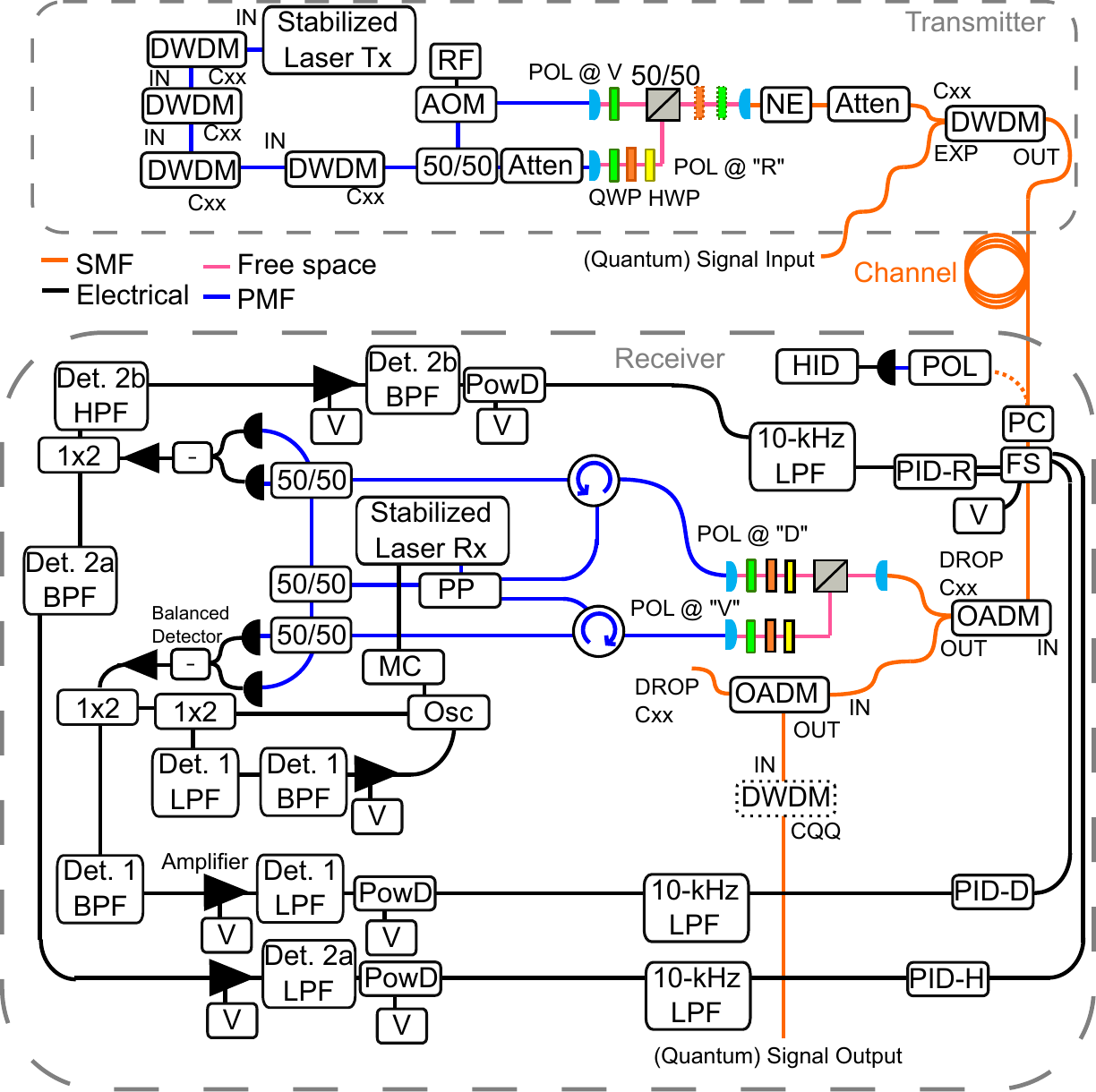}}
\caption{Automatic polarization compensation transmitter and receiver setup using laser references, wavelength multiplexing, and heterodyne detection. Elements with dashed borders are optional or temporary.  AOM: acousto-optic modulator. Atten : attenuator. BPF : band-pass filter. CQQ : {100-GHZ ITU DWDM channel number QQ using for the (quantum) signal}. Cxx : {100-GHZ ITU DWDM channel number xx used for the reference signals}. ``D'' : ``Diagonal'' polarization. Det. : detector. DWDM : dense wavelength-division multiplexing. EOM : electro-optic modulator. EXP : express port. FS : fiber squeezer. H : horizontal polarization. HID : human interface device. HWP : half-wave plate. LPF : low-pass filter. MC : microcomputer. OADM : optical add drop multiplexer. Osc : oscilloscope. PC : manual polarization controller. POL : polarizer. PowD : RF power detector. PP : Patch panel. QWP : quarter-wave plate. V : voltage supply.``V'' : ``Vertical'' polarization. {ITU : international telecommunication union.} The receiver waveplates with black border are motorized.}
\label{fig:apcsetup}
\end{figure}

{We now describe the system transformations for our implementation for which we have a flowchart (Fig.~\ref{fig:apcflowchart}) and experimental setup (Fig.~\ref{fig:apcsetup}); Experimental details and hardware particulars are described in \ref{app:APCdetails}. In the following, we treat the reference light classically. 

We start by preparing two reference signals from continuous-wave electro-magnetic waves represented here by their electric fields $\textbf{E}(f_{T},\textbf{R},t)$ and $\textbf{E}(f_{T}+X,\textbf{V},t)$ which are combined into the same spatial mode of a certain single-mode fiber, where $f_{T}$ is the frequency of the transmit laser, $X$ is the frequency shift from an acousto-optic modulator (AOM), $\textbf{R}$ and $\textbf{V}$ correspond to right-circular and vertical linear polarization vector representation, respectively, and $t$ is time. Propagating through the system to the heterodyne spectrometers affects the polarization:

\begin{figure}
    \centerline{\includegraphics[width=1\textwidth]{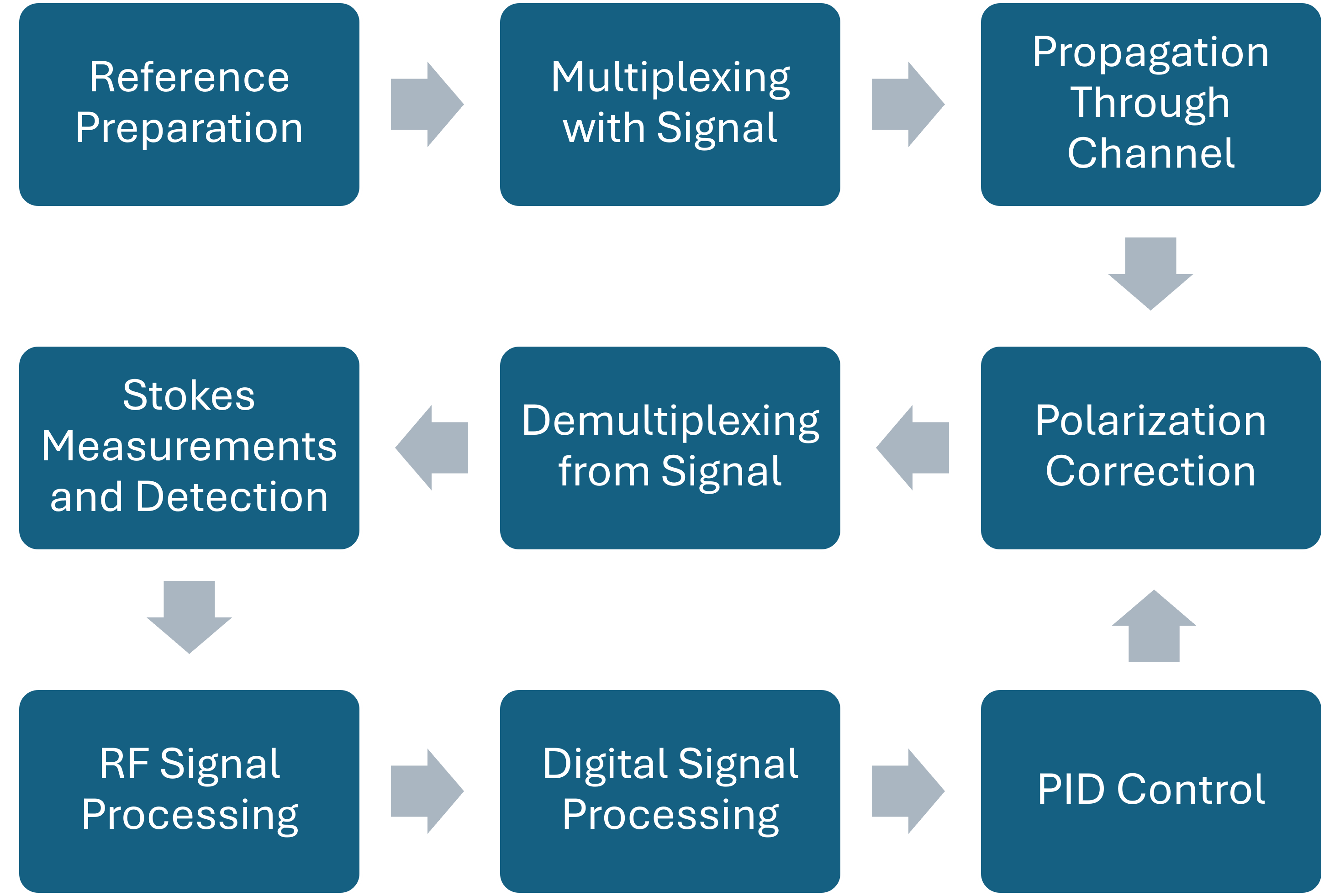}}
    \caption{{APC flowchart: The reference signals are prepared with filtering, frequency shifting, polarization, and attenuation. The reference signals are then multiplexed with the input signal and then all propagate through the channel. All signals then have polarization corrections applied before demultiplexing the input signal from the reference signals. The polarization and frequency of the reference signals is measured. RF and digital signal processing is used to select certain RF frequency bandwidths, apply amplification, filter noise, and measure the RF power. The PID takes the signal processing output and computes a polarization correction to apply which closes the feedback loop.}}
    \label{fig:apcflowchart}
\end{figure}

\begin{align}
{\big(E_V(f_{T},t) + E_V(f_{T}+X,t)\big)=}&\text{ }{\textbf{P} \textbf{U}_{\text{QWP1}}\textbf{U}_{\text{HWP1}}\textbf{U}_{\text{PCM}}\textbf{U}_{\text{FS}}\textbf{U}_{\text{Ch}}\big(\textbf{E}(f_{T},\textbf{R},t) + \textbf{E}(f_{T}+X,\textbf{V},t)\big)}\\
{\big(E_D(f_{T},t) + E_D(f_{T}+X,t)\big)=}&\text{ }{\textbf{P} \textbf{U}_{\text{QWP2}}\textbf{U}_{\text{HWP2}}\textbf{U}_{\text{PCM}}\textbf{U}_{\text{FS}}\textbf{U}_{\text{Ch}}\big(\textbf{E}(f_{T},\textbf{R},t) + \textbf{E}(f_{T}+X,\textbf{V},t)\big)},
\end{align}
}
{where the polarization transformations $\textbf{U}_{\text{Ch}}$, $\textbf{U}_{\text{FS}}$, and $\textbf{U}_{\text{PCM}}$ correspond to the channel, all fiber squeezers, propagation between correction (fiber squeezers) and measurement, respectively. $\textbf{U}_{\text{QWPn}}$ and $\textbf{U}_{\text{HWPn}}$ correspond to the quarter-wave plate and half-wave plates for the measurement connected to Detector $n$. $\textbf{P}$ is a polarizer aligned with the Heterodyne local oscillator polarization (slow-axis of the polarization-maintaining fiber). The $\textbf{U}_{\text{QWPn}}$ and $\textbf{U}_{\text{HWPn}}$ simultaneously serve multiple purposes: to correct for $\textbf{U}_{\text{PCM}}$, to serve as a quarter-wave plate for the R/L basis fiber squeezer, and to put the polarizer in the correct measurement basis. The calibration of these wave plates is described in \ref{app:APCproc}. Moreover $E_V$ and $E_D$ are the electric field components after the V and D polarization measurements heading to heterodyne spectrometer Detector 1 and 2, respectively.}

{
The heterodyne spectrometer local oscillator $E_{LO}(f_{R},t)$ also emits a continuous-wave electro-magnetic wave but at frequency $f_R$, where $Y=f_T - f_R$. After the heterodyne mixing and balanced photo-detection, each detector produces a difference photo-current with frequencies at $Y$ and $Y+X$ where the amplitude photo-currents is proportional to the input light's electric field amplitude due to the heterodyne method with some added noise from the LO shot noise. These photo-currents then go through a transimpedance amplifier (TIA) to produce an amplified (by frequency-dependent gain $G_{\text{TIA}}(f)$) voltage $v$ proportional to the photo-current $i$.

\begin{align}
v_V(Y,t) + v_V(X+Y,t)=&\text{ } i_V(Y) G_{\text{TIA}}(Y) + i_V(X+Y,t) G_{\text{TIA}}(X+Y,t)\\
v_D(Y,t) + v_D(X+Y,t)=&\text{ } i_D(Y) G_{\text{TIA}}(Y) + i_D(X+Y,t) G_{\text{TIA}}(X+Y,t).
\end{align}

{What we have described thus far can be interpreted as partial Stokes measurements on the reference signals, using a combination of polarization projections and heterodyne detection.} {These are partial Stokes measurements because only one polarization is measured per basis, which can be used as a Stokes measurement assuming the total power does not fluctuation substantially. After further signal processing, described below, these stokes measurements are used as PID inputs according to our algorithm explained above.}

At this point, the signals are split to supply the various feedback loops (three for polarization correction and one for frequency stabilization). We will focus on the signal transformations for the polarization correction loops; the frequency stabilization is discussed in \ref{app:APCdetails}. Each polarization correction control loop applies a correction by actuating a fiber squeezer (i.e., varying the phase of a VWP) along a certain basis state $s$; here we denote polarization measurements used as the input to the PID for rotations about basis state $s$ as $M_s$.

\begin{align}
M_D(t)=&10\log_{10}\bigg(\bigg<\big(F_{\text{L1}} G(Y) F_{\text{B1}} v_V(Y,t) + F_{\text{L1}} G(X+Y) F_{\text{B1}} v_V(X+Y,t)\big)^2\bigg>/50\bigg),\\
M_H(t)=&10\log_{10}\bigg(\bigg<\big(F_{\text{L2a}} G(Y) F_{\text{B2a}} v_D(Y,t) + F_{\text{L2a}} G(X+Y) F_{\text{B2a}} v_D(X+Y,t)\big)^2\bigg>/50\bigg),\\
M_R(t)=&10\log_{10}\bigg(\bigg<\big(F_{\text{B2b}} G(Y) F_{\text{H2b}} v_D(Y,t) + F_{\text{B2b}} G(X+Y) F_{\text{H2b}} v_D(X+Y,t)\big)^2\bigg>/50\bigg),
\end{align}

where $10\log_{10}\big(\big<x^2\big>/Z\big)$ denotes the power (dB) of $x$ dissipated into a load with impedance $Z$ as measured by our chosen power detector (Mini-circuits ZX47-60+). The logarithm from this detector is not ideal but can be tolerated by assuming there are small signal changes; which is true for polarization transformations within the system bandwidth once the loop is closed. $G(f)$ is the frequency-dependent gain of the RF amplifier (Mini-circuits ZKL-1R5+). $F_{\text{Bx}}$, $F_{\text{Lx}}$, and $F_{\text{Hx}}$ are the band-pass, low-pass, and high-pass filters, respectively, used for signal conditioning. See \ref{app:APCdetails} for details. These filters effectively enable each to focus on one of the reference signals for each measurement by significantly attenuating the other. Thus,

\begin{align}
M_D(t)&\approx\bigg<\big(F_{\text{L1}} G(Y) F_{\text{B1}} v_V(Y,t)\big)^2\bigg>,\\
M_H(t)&\approx\bigg<\big(F_{\text{L2a}} G(Y) F_{\text{B2a}} v_D(Y,t)\big)^2\bigg>,\\
M_R(t)&\approx\bigg<\big(F_{\text{B2b}} G(X+Y) F_{\text{H2b}} v_D(X+Y,t)\big)^2\bigg>.
\end{align}

Until this point, the electronics have been completely analog. But after the power detector, the signals are sampled and digital signal processing is applied. Each power detector output goes through a digital 8th-order 10-kHz Butterworth low-pass filter then into the digital PID module as described above. The output of the PID modules are converted back to analog and sent to the fiber squeezer driver/amplifier analog input. $M_D$ is used as the input to PID-D which controls a fiber squeezer oriented at $45^{\circ}$. Likewise, $M_H$ is used as the input to PID-H which controls a fiber squeezer oriented at $0^{\circ}$. Finally, $M_R$ is used as the input to PID-R which controls a fiber squeezer oriented at $45^{\circ}$ that is rotated into the R/L basis by a DC calibration voltage applied to proceeding fiber squeezer. 

}
With the equipment set up and algorithm established, calibration of the receiver reference measurement bases and PID settings remain. Then the APC can be enabled. Details on those procedures are in \ref{app:APCproc}. For the following tests and demonstrations, we used the base algorithm described above without the slope sign included for a proof-of-principle characterization of the overall algorithm principles.

\subsection{Entangled Photon Source and Tomography}
As shown in~Fig.~\ref{fig:ChattQLAN}, the source design is based on fiber-based Sagnac loop~\cite{Li_2005source,Fan_2007source,Vergyris_2017,Alshowkan_2022}. For portability, a 785-nm fiber-pigtailed laser (Thorlabs) hosted on a laser diode driver (mw-tech) followed by an optical isolator is amplified using a semiconductor optical amplifier (SOA, AERO Diode) pumps a fiber-coupled 12-mm-long periodically poled lithium niobate (PPLN) ridge waveguide (AdvR) for type-0 phase-matched spontaneous parametric downconversion (SPDC). The loop is designed for generating spectrally correlated, polarization-entangled photons in the ideal Bell state $\ket{ \Phi^+}\propto\ket{HH}+\ket{VV}$. The resulting bandwidth spans approximately 18~THz, covering the entire C- and L-bands~\cite{Alshowkan_2022}. The SOA is connected to a half-waveplate liquid-crystal variable retarder (Thorlabs) followed by a 780/1550-nm {wavelength-division multiplexer (WDM)} which passes the light to a fiber-based polarizing-beam splitter/combiner (Thorlabs) to coherently pump the waveguide from both directions and then combines the generated bandwidth. The returned bi-photons to the 780/1550-nm WDM get reflected to a C/L-band WDM (ACPhotonics) where the higher (lower) frequencies are routed to a C-band (L-band) wavelength-selective switch (WSS, Finisar). Each WSS enables carving the bandwidth with 6.25~GHz resolution that is aligned with ITU grid (ITU-T Rec. G.694.1) to numerous output fibers. For stabilization against the environmentally induced fluctuations in splitting ratio previously observed in~\cite{Alshowkan_2022}, we tap 1\% of the light in each direction to feedback the LCVR to maintain the pump splitting ratio~\cite{Lu_2023}. In our setup we enable the WSSs to output a 100~GHz bin frequency-correlated channel. 
We use quantum state tomography to characterize the quality of polarization entanglement using a pair of polarization analyzers each consisting of a quarter-wave plate (QWP; Thorlabs), a half-wave plate (HWP; Thorlabs), and a polarizing-beam splitter (PBS; Thorlabs), where the wave plates are operated by a set of motorized rotation mounts (K10CR1; Thorlabs). The detection pulses from the detectors are timestamped by a field-programmable gate array (FPGA; RFSoC 4$\times$2) time-to-digital converter (TDC), synchronized using White Rabbit node (WR; Safran) timing system as described in~\cite{Alshowkan_2021, Alshowkan_2022jocn}. An upgrade from our previous design in~\cite{Alshowkan_2022jocn}, here the FPGA-based TDC directly accepts the WR 10-MHz clock---overcoming the need for frequency doubling. Finally, the measurement and coincidence analysis systems are outlined in~\cite{IDref18, alshowkan_2024}. By applying a pre-calibrated time delay to account for the varying path lengths photons travel to reach each detector and using about a 660~ps coincidence window, we collect coincidences for each measurement setting between pairs of nodes. These procedures are repeated for 36 polarization projections across rectilinear (H/V), diagonal (D/A), and circular (R/L) bases forming an over-complete dataset.
\subsection{Metropolitan Quantum Network}
\label{sec:methodsEPB}

\begin{figure}
\centerline{\includegraphics[width=1\textwidth]{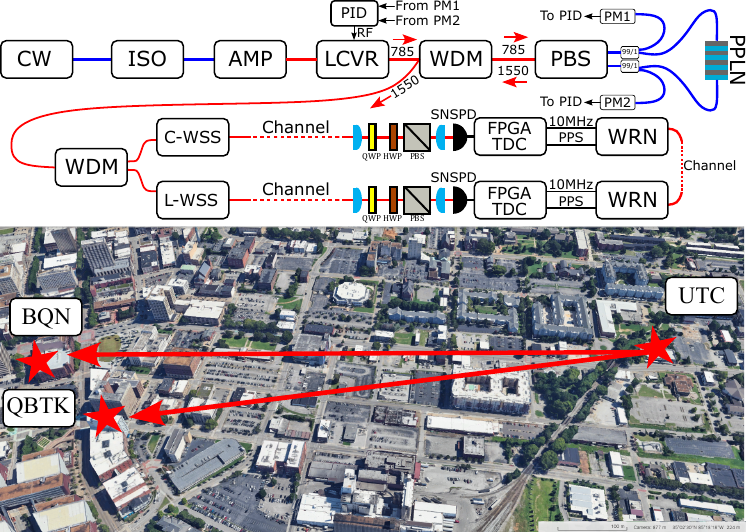}}
\caption{Entanglement source and polarization analyzer setup (upper inset). Map of the entanglement distribution stabilized with APC on a metropolitan quantum network (lower inset) with the source at University of Tennessee--Chattanooga, L-band photons going to Broad St. quantum node (BQN), and C-band photons going to Qubitekk quantum node (QBTK).
10-MHz: a reference clock.
AMP: optical amplifier.
CW: continuous-wave laser.
FPGA: field-programmable gate array.
HWP: half-wave plate.
ISO: optical isolator.
LCVR: liquid crystal variable retarder.
PBS: polarizing-beam splitter.
PID: proportional-integral-derivative controller.
PM: optical power meter.
PPLN: periodically poled lithium niobate.
PPS: pulse-per-second.
QWP: quarter-wave plate.
SNSPD: superconducting nanowire single-photon detector.
TDC: time-to-digital converter.
WDM: wavelength division multiplexer.
WRN: White Rabbit node.
WSS: wavelength-selective switch.
Black lines: electrical cables.
Blue lines: polarization maintaining fiber.
Red lines: non-polarization maintaining fiber.
Dotted lines: a communication channel which could be a fiber spool or deployed fiber depends on the setup.
}
\label{fig:ChattQLAN}
\end{figure}

Here we describe the metropolitan quantum network used in testing and the main configuration used for entanglement distribution [Fig.~\ref{fig:ChattQLAN} lower inset]. In this configuration, we place the entanglement source with both L-band and C-band APC transmitters at the University-of-Tennessee-Chattanooga (UTC) node on the Electronic Power Board (EPB) quantum network. We then distribute the L-band photons through an APC transmitter then 3.2~km in fiber to a node at the Broad-St. quantum hub (BQN) which houses an APC receiver, polarization analyzer, and single-photon detectors. Similarly, we distribute the C-band photons through an APC transmitter then 3.5~km to the Qubitekk quantum node (QBTK), housing an APC receiver and polarization analyzer. The C-band photons were then transferred 1.3~km to the 10th-St. quantum hub for single-photon detection.

In this configuration, the losses for the C-band (L-band) path after the source include -0.5 from C/L-band wavelength-division multiplexer, -5~dB from WSS, -0.5~dB from APC transmitter, -5.5~dB (from propagation from UTC to Qubitekk node and same for UTC to Broad St. node), -1.3~dB (-2~dB) from APC C-band (L-band) receiver, -4~dB (-2.5~dB) from polarization analyzer, -0.5~dB from additional {100-GHz dense-wavelength-division multiplexer (DWDM)}, -6.8~dB (-7.2~dB) path from propagation to detectors, and estimated -1~dB from superconducting nanowire single-photon detectors (Quantum Opus) for a total C-band (L-band) transmission loss of -19.6~dB (-19.2~dB).

\section{Results}
\label{sec:results}
\subsection{Classical polarization drift characterization}
In preparation for testing the APC on deployed in-ground fiber on the EPB quantum network in Chattanooga, Tennessee, USA, we characterized the bandwidth of the polarization fluctuations over a 5-day period. Essentially, we transmitted a polarized laser from the Qubitekk quantum node to the node at Broad-St. quantum hub via the node at the UTC to pass through both links planned for the entanglement distribution test. The received polarization was measured using a fiber polarization beamsplitter and power meters. The analog power meter outputs were sampled then fast Fourier transforms (FFT) were calculated for various sample rates (ranging from 1.5~mHz to 100~MHz) at various intervals continuously for 5 days. More information about this characterization configuration is in \ref{app:polchar}.

\begin{figure}
\centerline{\includegraphics[width=1\textwidth]{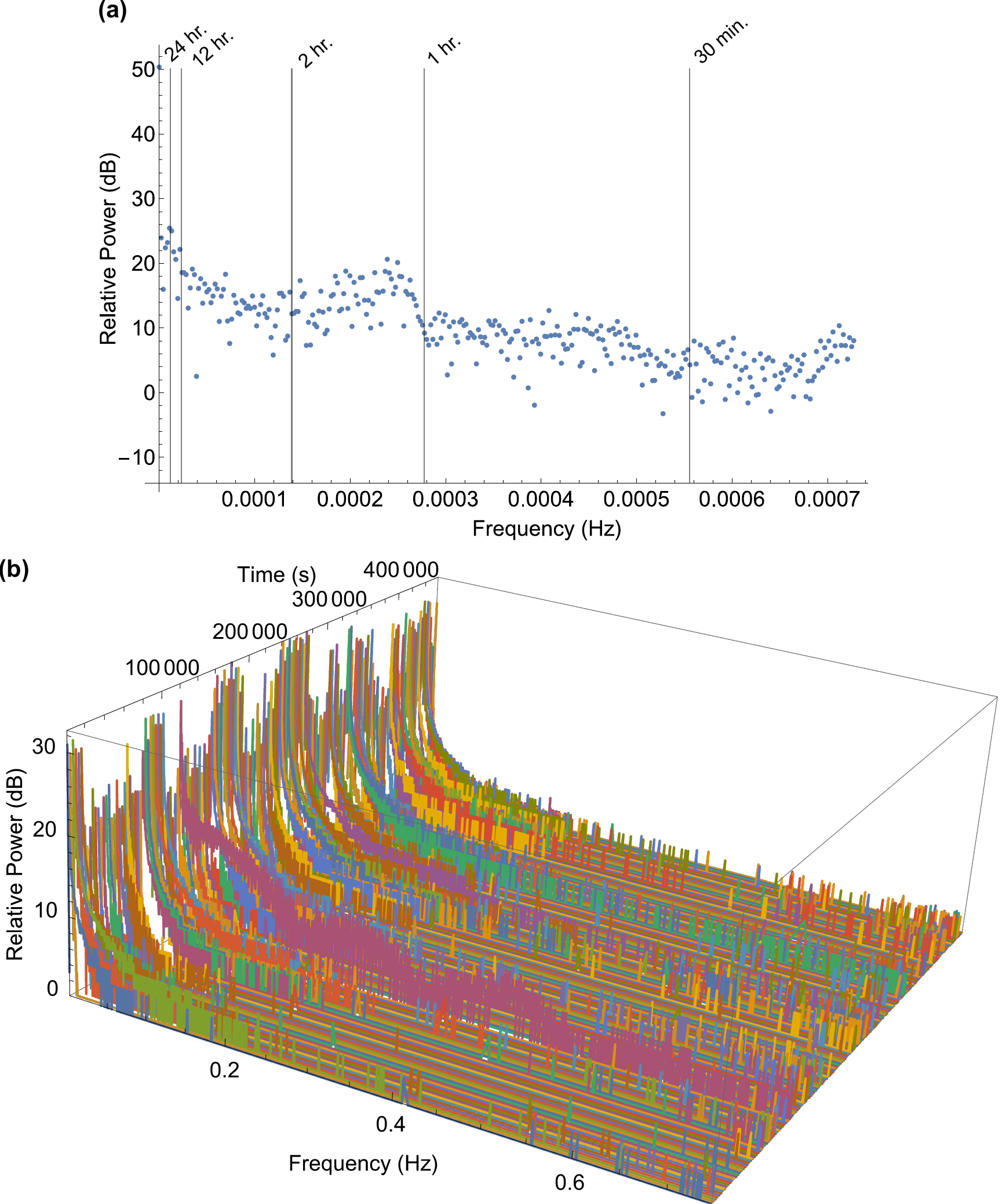}}
\caption{Polarization drift characterization. (a) Fast Fourier transform (FFT) of 670 1.46-mHz power samples acquired over 5 days showing polarization variation on the EPB quantum network. (b) FFTs of polarization variation from 4096 power samples collected at 1.49-Hz sample rate. The FFTs are repeated continuously for 5 days as 4096 samples are collected (about every 46 min.).}
\label{fig:polcharvert}
\end{figure}

Over 5 days, we collected 670 samples at 1.46-mHz sample rate from which the FFT [Fig.~\ref{fig:polcharvert}(a)] reveals information about long-term drifts. We find peaks on time scales of 1 and 24 hrs with the drift being noticeably larger on time scales > 1 hr. compared to < 1 hr. Nonetheless FFTs in Fig.~\ref{fig:polcharvert}(b), which repeated about every 46 minutes (each collecting 4096 1.5-Hz samples) for 5 days, show there is a variable amount of polarization noise in the 10 - 100 mHz frequency range as well. At higher frequencies, we found there is a small amount of even more rare (a few times over 5 days) noise occasionally between 1 - 400 Hz.

\subsection{APC characterization with classical test signals}
To demonstrate the control bandwidth of the APC, we induced polarization drift using an external fiber squeezer device before the APC receiver. In this configuration, we use our APC set up for C-band operation; an L-band APC will be tested below. A test signal laser with wavelength 1559.75~nm was multiplexed with the C-band APC using reference wavelength 1558.2~nm and the two signals propagated together through a 3.5-km fiber across Chattanooga from the UTC to the Qubitekk quantum node on the EPB quantum network. After the signals reached the Qubitekk quantum node, controlled polarization drift was induced using an fiber{-}squeezer device, then the test signal was split 50/50 using a fiber fused coupler where half went to a polarizer to be detected by a power meter (Thorlabs PM101A, S154C) while the other half went through the APC receiver for correction and then went to a polarizer to be detected by a power meter (Thorlabs PM101A, S154C). Both power meter analog outputs were then recorded with data acquisition hardware ({Liquid Instruments} Moku:Go).

\begin{figure}
\centerline{\includegraphics[width=1\textwidth]{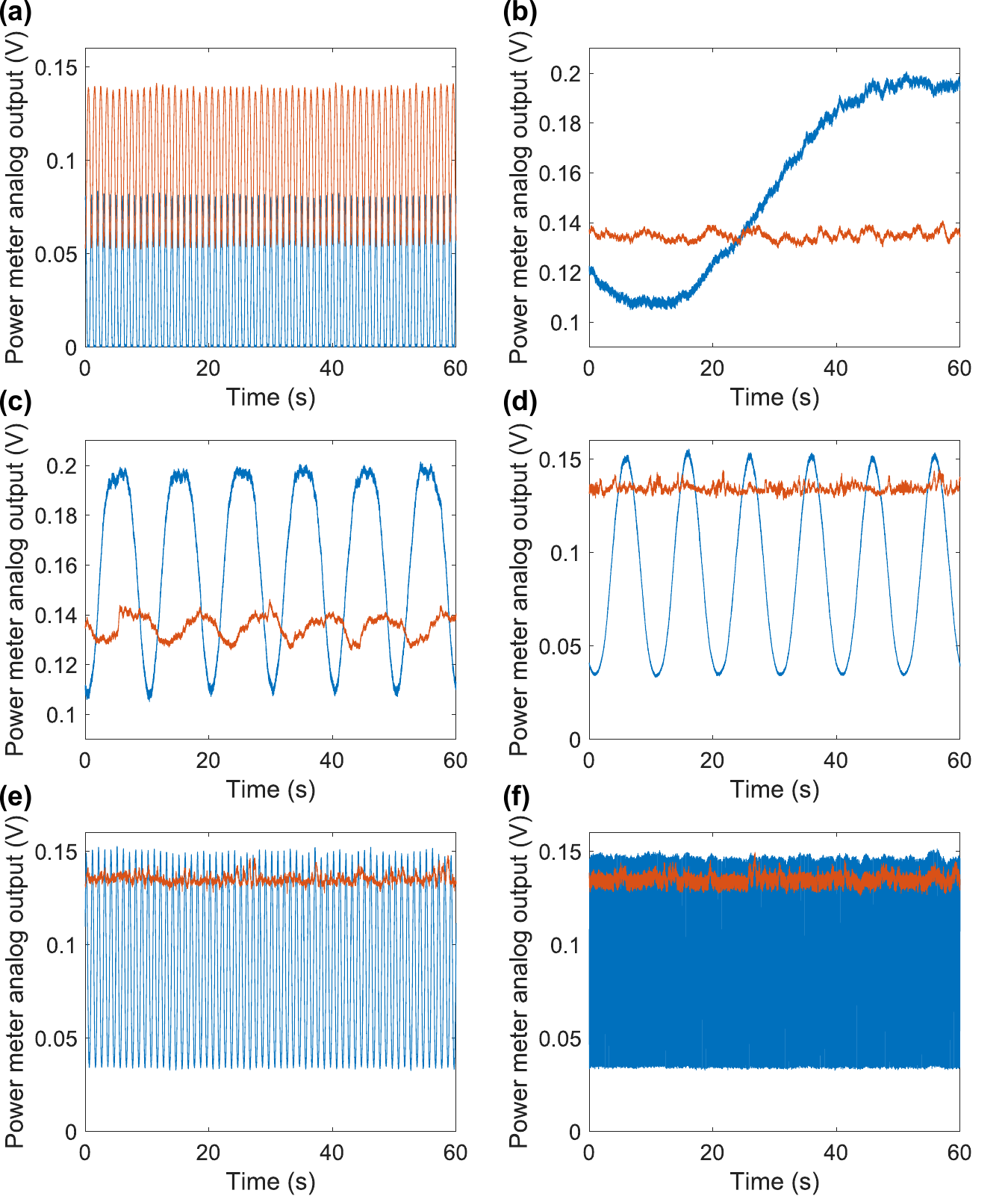}}
\caption{C-band Classical test signals with(out) C-band APC. APC reference laser wavelength was 1558.2 nm. Test signal wavelength was 1559.75 nm, about 2 100-GHz channels away.  (a) Test signals after 1-Hz induced polarization drift without APC (blue) and through disabled APC (orange). Test signals after 10-mHz (b) or 100-mHz (c) induced polarization drift without APC (blue) and with APC (orange) using about 1-Hz PID integration bandwidth. Test signals after 100-mHz (d), 1-Hz (e), or 10-Hz(f) induced polarization drift without APC (blue) and with APC (orange) using about 1-kHz PID integration bandwidth.}
\label{fig:CbFSFGtests}
\end{figure}

Fig.~\ref{fig:CbFSFGtests}(a) provides calibration of a characteristic oscillation range for both test signals (with the APC off in this case). After engaging the APC to correct for drift, in Fig.~\ref{fig:CbFSFGtests}(b), we see the APC with 1-Hz PID integration bandwidth tracks a 10-mHz polarization drift without issue, though there is slightly increased polarization noise on the APC-corrected test signal compared to the non-APC test signal. This increased polarization noise is somewhat due to the inherent nature of close-loop control which will necessarily have at least slightly higher noise compared a stable uncontrolled signal. Additionally, there is excess polarization noise due to the imperfect frequency stabilization. With a 1-Hz PID integration bandwidth, most of the residual frequency-induced power variation is averaged out but not all is averaged out leading to increased polarization noise. This is more pronounced in the C-band testing where the reference and LO APC lasers did not have the manufacturer's no-drift frequency calibration which in the L-band APC lead to lower excess polarization noise. In Fig.~\ref{fig:CbFSFGtests}(c), we see the 1-Hz PID integration bandwidth is not able to fully track at 100-mHz variation though there is still significant correction applied compared to the uncorrected signal. Empirically, we have found that for complete tracking, the drift should be about 100 times less than the integration bandwidth (the integrator gain is variable with the frequency, lower at higher frequencies).

This 1-Hz control bandwidth is sufficient in most cases for few kilometer lengths of in-ground fiber but if higher-speed drift is expected, the integration bandwidth can be increased, at the cost of transferring more of the frequency-induced power variation into polarization noise onto the test signal in our current implementation where the frequency stabilization has room for improvement. In Fig.~\ref{fig:CbFSFGtests}(d)-(f), using a 1-kHz PID integration bandwidth, we apply 100-mHz, 1-Hz, and 10-Hz induced polarization variation. These tests show the APC is able to track these higher speed signals completely with only slight increase in polarization noise starting to show for 10-Hz drift. In \ref{app:addclasstest}, we provide further testing up to 100-Hz which has similar results to Fig.~\ref{fig:CbFSFGtests}(c) in the 1-Hz PID integration bandwidth case.

\begin{figure}
\centerline{\includegraphics[width=1\textwidth]{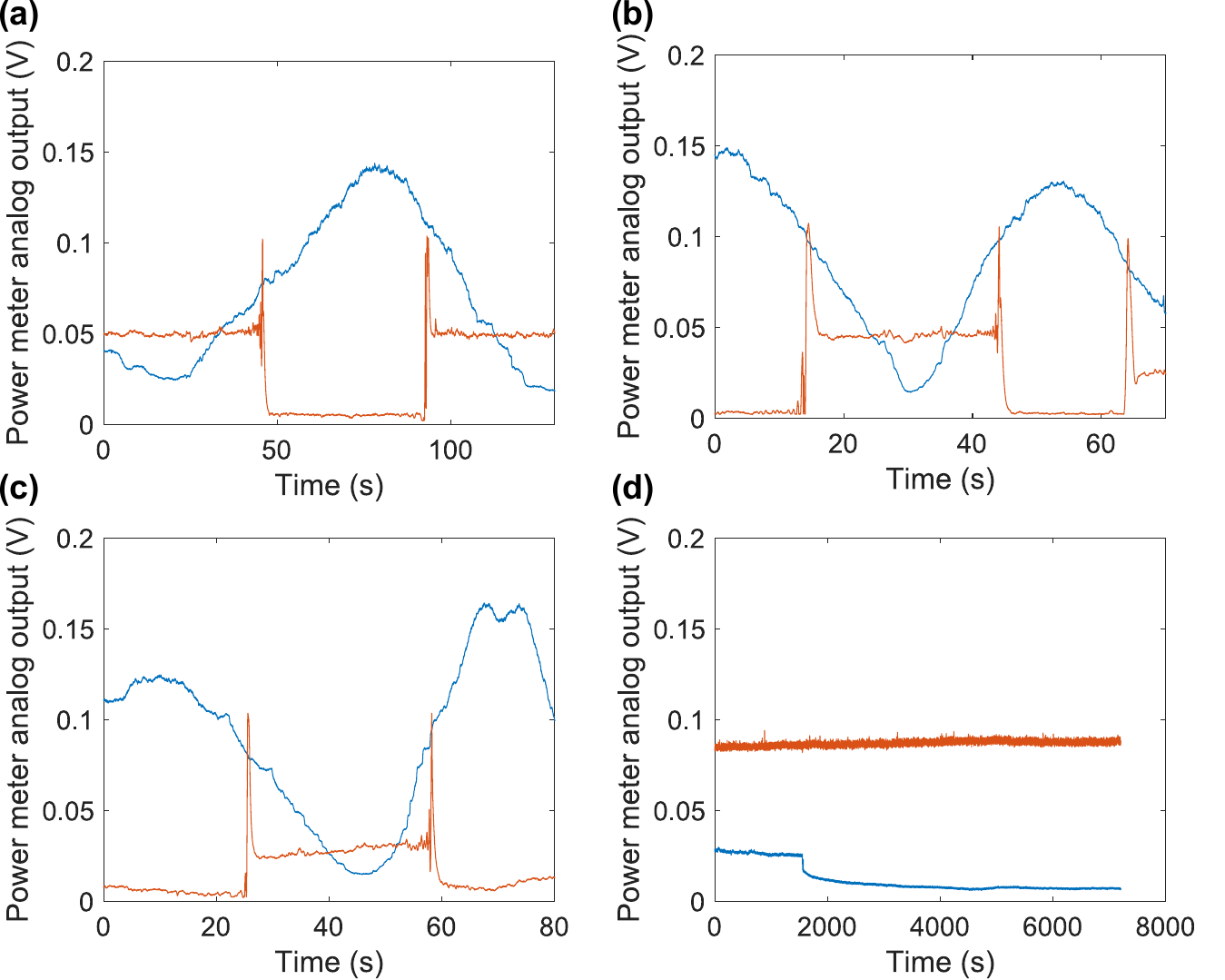}}
\caption{L-band Classical test signals with(out) L-band APC. APC reference laser wavelength was 1581.2 nm. 1579.6-nm (a), 1577.2-nm (b), and 1573.2-nm (c) test signals after manual induced polarization drift bypassing APC (blue) and through APC (orange) using about 1-Hz PID integration bandwidth. (d) 1579.6-nm test signals after natural drift across metropolitan Chattanooga bypassing APC (blue) and through APC (orange) using about 1-Hz PID integration bandwidth.}
\label{fig:Lbtests}
\end{figure}

Furthermore, we also set up a test configuration for the L-band APC. In this configuration, a test signal laser with initial wavelength 1579.6~nm was multiplexed with the APC using reference wavelength 1581.2~nm and the two signals propagated together through a 3.2-km fiber across Chattanooga, TN from the UTC to a node at the Broad-St. quantum hub on the EPB quantum network. After the signals reached the Qubitekk quantum node, controlled polarization drift was induced using a manual polarization controller. The rest of the test configuration is identical to the above. In Fig.~\ref{fig:Lbtests}(a) the polarization of the 1579.6~nm test signal (2 100-GHz channels away from 1581.2~nm) is held stable while the PID control is within range. Here is an example of the limited control range of the base PID algorithm which does not include the slope sign. Depending on where the fiber squeezers are in their range when the jump to a new fringe happens, the control can continue on a new place or may jump out of range. At slower integration bandwidths, empirically we have found it more likely to resume control at the next fringe, whereas for higher integration bandwidths the integral windup usually carries the PID output to the edge of the range.

Next, the test signal is reconfigured to 1577.2~nm (5 channels away) in Fig.~\ref{fig:Lbtests}(b) and there is now a slight tilt to the controlled signal which is more pronounced in Fig.~\ref{fig:Lbtests}(c) where the test signal was again reconfigured (to 1573.2~nm, 10 channels away). This provides a partial characterization of the polarization-mode dispersion sensitivity of the device. Using longer fibers and having the polarization drift either before or throughout propagation would show a more pronounced affect. But in the case of about 10{-}km length fibers, the 2-channel spacing is shown to be sufficiently insensitive to polarization-mode dispersion. This is further evidenced by Fig.~\ref{fig:Lbtests}(d), where the APC corrects for the natural drift of the fiber across metropolitan Chattanooga over the course of several hours. The residual drift of the test signal corrected by the APC can be  attributed to slight calibration drift of the measurements{,} which we have observed to be about 3{$-5^{\circ}$} within about 24 hours{, and to residual effects of polarization-mode dispersion}. The classical tests shown thus far have been helpful to ascertain the bandwidth performance of the APC but, using a polarizer, they have not been a complete characterization of the polarization channel stability. In the following subsections, we will show more complete polarization characterization, including quantum process and state tomography.

\subsection{Entanglement-assisted process tomography of APC-stabilized fiber}

To characterize the stability provided by the APC on a complete polarization channel (as opposed to the stability of a single SOP) and to simultaneously characterize its fitness for use in quantum communication, we employ entanglement-assisted process tomography, a special case of ancilla-assisted process tomography~\cite{PhysRevLett.90.193601}, specifically  we use the analysis developed in Ref.~\cite{lukensEAPT}. In traditional quantum process tomography~\cite{doi:10.1080/09500349708231894,PhysRevLett.78.390}, multiple different quantum states traverse the process under test and are each characterized by quantum state tomography. The results of which are aggregated to estimate what is the quantum process under test. Using the nature of two-qubit entanglement enables measurements of a two-qubit tomography, measured after one of the qubits traverse the process, to be reanalyzed to estimate the quantum process traversed by one of the entangled qubits.

For this characterization of the APC, it is not of concern what is the exact process but that it remains stable over time. Thus, we will take a series of tomographies after enabling the APC and compare the estimated processes $\mathfrak{C}_n$ (Choi matrices) using the process fidelity. We calculate the relative process fidelity (RPT) of $\mathfrak{C}_1$ and $\mathfrak{C}_2$ using the generalized Ulhmann fidelity for any two positive semi-definite operators~\cite{Wilde}:
\begin{equation}
F_P(\mathfrak{C}_1,\mathfrak{C}_2)=\Bigg(\Tr \sqrt{\sqrt{\mathfrak{C}_1}\mathfrak{C}_2\sqrt{\mathfrak{C}_1}}\Bigg)^2.
\end{equation}

For the first two-qubit tomography measured in a series, we will use as the required reference tomography in the analysis of Ref.~\cite{lukensEAPT} to estimate state imperfections. Using that reference tomography, we can analyze the second measured tomography providing the first estimated process. From there, we can continue to measure tomographies and estimate the processes from them. 

Prior to any quantum signal using the APC, we measured the noise from insufficient isolation filtering and spontaneous Raman scattering (sRs) generation. {To characterize the remaining reference light after the isolation filtering, we set up the} APC transmitter and receiver back{-}to{-}back {with a 2-m patch cord to minimize spontaneous Raman scattering from the channel between them. In that configuration,} we measured the photon counts emitted from the quantum-signal output port of the receiver after the two OADMs in the APC (acting as notch filters to remove the reference light). The photon counts were detected by superconducting nanowire single-photon detectors (Quantum Opus One) and the counts were integrated by a counter (Keysight MSOX4104A). We did not measure any counts above the detector dark counts (about 200 cps). {This implies that the noise from the reference signals themselves getting into the (quantum) signal due to insufficient filter isolation is 
 so low it is below the noise of our single-photon detectors.} With the received reference power, after a 5-km fiber spool, measured at -50 dBm before the APC receiver, we measured 20 kcps of sRs emitted from the quantum-signal output port. Due to the notch filtering used, those counts come from sRs photons at wavelengths across the C- and L-band and further. In certain situations, that could be an intolerable amount of noise depending on the coincidence-to-accidental ratio. In those cases, further filtering can be added. When we added a 100-GHz DWDM, the filtering was sufficient that we did not measure any counts above the detector dark counts. We decided to use the additional 100-GHz DWDM for the quantum tests in this work; thus, in this configuration, the coexistence of the classical APC references and the quantum signal added no appreciable noise.

To characterize the APC using entanglement-assisted process tomography, we transmit a 100-GHz channel from the entanglement source, centered on 1579.6~nm, through L-band APC transmitter (with references at 1581.2~nm) then through a 5-km spool heated by variable electronics waste heat (computer fan) followed by the APC receiver and then to the polarization analyzer for tomography measurement. The other corresponding entangled photons (a 100-GHz channel, centered on 1559.8~nm) were measured by the polarization analyzer directly after the WSS as a reference. This is diagramatically shown in Fig.~\ref{fig:ChattQLAN} where the L-band channel is encapsulated by an APC transmitter and receiver and the C-band channel is a short patch cord.

\begin{figure}
\centerline{\includegraphics[width=1\textwidth]{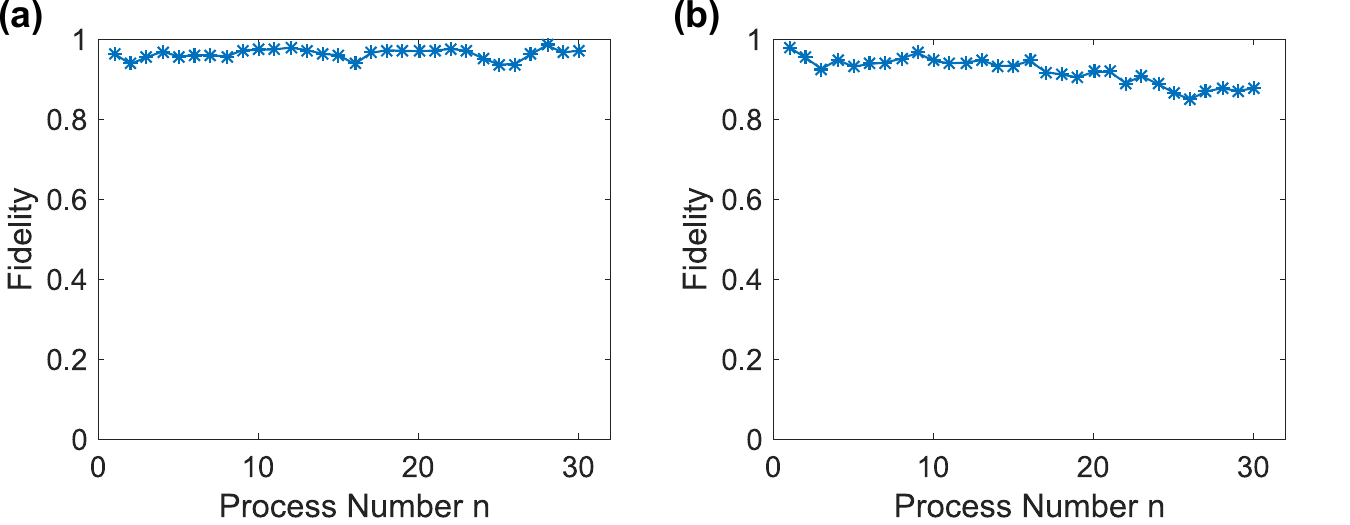}}
\caption{Relative process fidelities (RPF) for two successive series of process tomographies. (a) RPF between $\mathfrak{C}_n$ and $\mathfrak{C}_{n+1}$ and (b) RPF between  $\mathfrak{C}_1$ and $\mathfrak{C}_n$ with the APC enabled for the L-band photon going through a 5-km spool. Integration time per measurement setting was 2~s using a 660-ps coincidence window, collecting an average  $3.9x10^4$ coincidence counts per 36-setting tomography.}
\label{fig:EAPTresults}
\end{figure}

In this configuration, we measured a series of tomographies with the APC enabled. For tomographies measured with the APC enabled, the RPT between $\mathfrak{C}_n$ and $\mathfrak{C}_{n+1}$ are shown in Fig.~\ref{fig:EAPTresults}(a) and between $\mathfrak{C}_1$ and $\mathfrak{C}_n$ are shown in Fig.~\ref{fig:EAPTresults}(b). Here we find RPT stabilized through both comparisons due to the action of the APC stabilizing the complete polarization channel. For successive tomographies the measured average relative process fidelity is $0.96\pm0.01$ measured over 4.5 hrs. Moreover, the measured average relative process fidelity between $\mathfrak{C}_1$ and $\mathfrak{C}_n$ is $0.94\pm0.03$. This decrease (increase) in fidelity average (standard deviation) and the slight tilt in Fig.~\ref{fig:EAPTresults}(b) we attribute to calibration drift from our passively temperature-stabilized measurement system {and} there is a contribution from polarization-mode dispersion as well. {Comparing the calibrations done successive days, shows about $3-5^{\circ}$ shift in the optimal waveplate angles from which we estimate a few percent error due to the calibration drift. The rest is likely due to polarization-mode dispersion.} See \ref{app:partEAPTcomp} for a comparison without the APC enabled providing an upper bound to the fidelity variation that could have been observed.

\subsection{Entanglement distribution using APC across metropolitan network}
Having demonstrated the functionality of the APC at bandwidths relevant to distribution through in-ground fiber, we move on to using the APC for long-term stabilized polarization entanglement distribution on a metropolitan quantum network as described in Sec.~\ref{sec:methodsEPB}. In this specific case, the WSS is configured to transmit a 100-GHz channel from the entanglement source, centered on 1579.6~nm, to the L-band APC transmitter (with references at 1581.2~nm) and corresponding entangled photons (another 100-GHz channel, centered on 1559.8~nm) to the C-band APC transmitter {(with references at 1558.2~nm)}. Each are then distributed across the network receiving the APC correction after transmission.

After enabling each APC, a series of two-qubit polarization tomographies were repeatedly measured one after another and analyzed~\cite{Lukens2020,Lu2022b}. The APC operated continuously---correcting the polarization for over 30 hours---until one of the $\pi$-range limits was hit then the APC relocked to another fringe. As mentioned in Sec.~\ref{sec:methodsAPC}, this is a limitation with an algorithmic fix though we did not yet implement this additional complication for these tests. During this period, we collected 95 successive two-qubit tomographies (each taking about 16-17 minutes). 

\begin{figure}
\centerline{\includegraphics[width=1\textwidth]{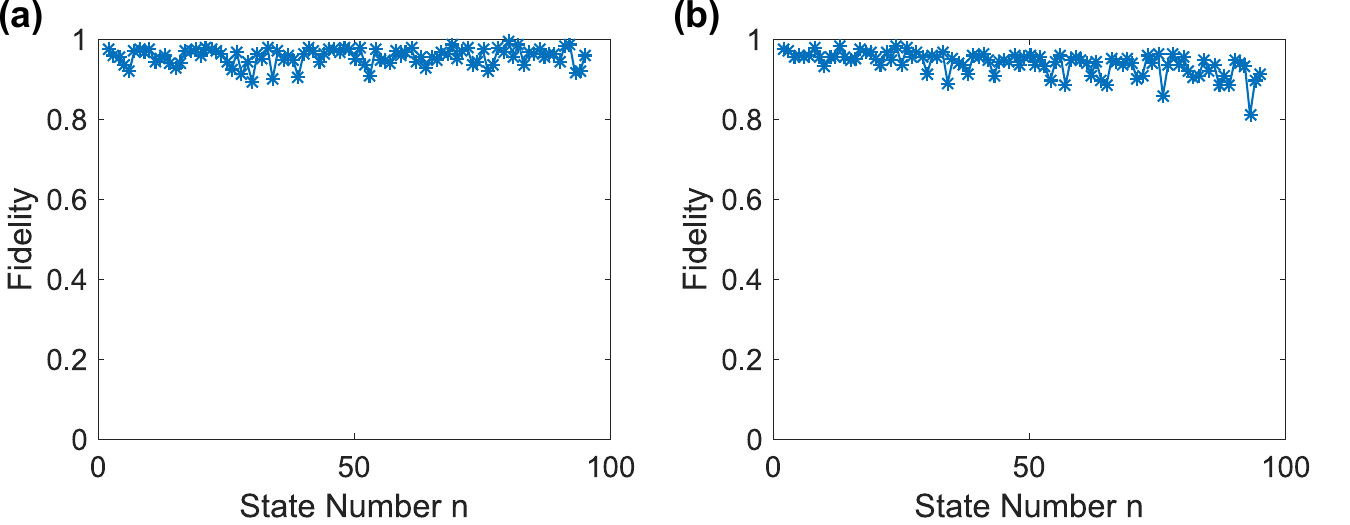}}
\caption{Relative two-qubit state fidelities for entanglement distribution with APC correction over a metropolitan quantum network (a) between successive estimated states (i.e., $\rho_n$ and $\rho_{n+1}$), and (b) between the first estimated state and successive states (i.e., $\rho_1$ and $\rho_{n}$). Integration time per measurement setting was 15~s using a 660-ps coincidence window, collecting an average 310 coincidence counts per 36-setting tomography.}
\label{fig:EntDistdata}
\end{figure}

In Fig.~\ref{fig:EntDistdata}, we show the relative generalized Ulhmann fidelity for successive estimated two-qubit states and the relative fidelity between the first estimated state and successive states. Due to the low coincidence rate, there is noticeable noise from statistical noise. Despite that, Fig.~\ref{fig:EntDistdata}(a), shows a flat trend over the >30-hour period with a average relative fidelity of $0.96\pm0.02$ between successive estimated states implying the APC managed the stability on the order of two tomographies throughout this period. From logged data of the APC PID outputs, we have estimates of the fiber polarization drift present when the APC was enabled (See \ref{app:PIDoutentdist}). There was drift on the time scale of two tomographies but not necessarily large enough to be seen in this graph with the statistical noise present. Thus, Fig.~\ref{fig:EntDistdata}(a) primarily provides validation of two independent APC inherent stability on the time scale of a half hour over 30 hours. 

Fig.~\ref{fig:EntDistdata}(b) on the other hand shows the relative fidelity between the first estimate state and successive states. On this time scale ($>30$ hrs), we did measure significant polarization drift corrected by the APC outputs. The average relative fidelity in this case is $0.94\pm0.03$. The decrease (increase) in average (standard deviation) of relative fidelity can be seen graphically by the slight trend down in Fig.~\ref{fig:EntDistdata}(b). There may also be an increase in standard deviation over time but it is uncertain whether that is due to increased sample numbers or an actual change in the standard deviation over time. Overall, the fidelity is held stably by two independent APC for entanglement distribution for over 30 continuous hours without interruption.

\section{Discussion}
\label{sec:discuss}
As quantum networks continue to mature and increase in technology readiness level, they will be used in more real-world conditions and less controlled lab-like environments. One of the first issues encountered in these deployments is polarization drift. For discrete-variable quantum information which is often encoded on the polarization of single-photons, the methods presented here provide solutions to balance the somewhat contradicting demands of 100\% up-time, high-bandwidth detection and correction, low-insertion loss, and negligible added noise.  Here we leverage heterodyne detection to enable the benefits of using high-power laser reference signals (which include low-noise high-bandwidth error signals) while maintaining the benefits of dim reference signals detected by single-photon detectors (i.e., dim signals do not produce much Raman or cross-talk noise) all the while enabling 100\% up-time using {wavelength}-multiplexing. {In addition to the discrete-variable/single-photon application we demonstrate, t}hese methods can also be use for continuous-variable experiments using homodyne detection or in classical networking {(with direct or coherent detection)} where the requirements  for  dim reference signals are much more relaxed~\cite{IDref20}. 

There are several benefits to automated polarization compensation using this method compared to other demonstrated methods. Using a {wavelength}-multiplexed attenuated laser source as the reference signals provides an ever-present reference signal. This signal can be attenuated so there is not significant Raman or cross-talk noise in the quantum signal after filtering with moderate insertion loss. By using heterodyne detection, the best of both worlds is achieved, namely, ultra-high-bandwidth (up to GHz have been demonstrated) and ultra-sensitive detection near the single-photon level (per mode). Moreover, the local oscillator and reference lasers do not need to be phase coherent which simplifies the overall setup. The local oscillator used in heterodyne detection is a bright mW laser which enables the use of conventional fast photodiodes in the balanced detector within the homodyne detector. Using fast photodiodes and modern high-speed electronic amplifiers can enable a balanced detector with GHz bandwidth. Moreover, because heterodyne detection can detect down to about one input photon per mode, using a narrowband local oscillator enables the detection of signals with total power below -100 dBm where the spontaneous Raman scattering generated from such a signal is negligible in most instances and the cross-talk can most likely be reduced to tolerable, even negligible, levels with reduced demultiplexing filtering. 

Taking advantage of the high-speed detection signals allows high-bandwidth polarization actuators to be used like the piezo fiber squeezers or the lithium niobate modulators which should enable stabilization in the most trying of conditions, e.g., aerial fiber on a stormy day. It may be helpful to cascade the fiber squeezers with a faster lithium niobate modulator for enhanced control and to enable control over larger phase shifts since the lithium niobate phase range is limited. Appropriate PID tuning while increasing the closed-loop PID bandwidth should enable use of these higher-bandwidth actuators. Precise control at higher speed would also benefit from improved relative frequency stabilization for the heterodyne measurements; this improved stabilization is possible with current technology and further engineering as discussed in {\ref{app:APCdetails}}.

Moreover, using the three independent PID loops for control provides speed and simplicity which should lead to less downtime when other algorithms would need to search to find the setpoint again after a large disturbance. Though if the axes of the variable wave-plates are not set correctly, their great circles will not be orthogonal at R leading to reduced control. Currently, the algorithms are limited by the phase-shift range of the variable wave-plates but so-called endless control has been shown with multiple types of polarization controllers, including variable wave-plates. This could possibly be augmented to the proposed control method, possibly using the switching retarder of  Ref.~\cite{Martinelli2006}. Though it appears that endless control with limited retardance actuators is not possible for polarization channel stabilization compared to SOP stabilization since in SOP stabilization one can rotate about the SOP axis without changing the stabilized SOP. This sort of transformation does not appear possible for channel stabilization since a polarization channel includes any applied polarization transformation.

The entanglement distribution results show some slight drift {over} longer periods of time. The current implementation used passive temperature stabilization in the transmitter (vented boxes) and receiver (vented boxes with foam{-}insulated fiber) and the leftover temperature drift is likely {a} main contributor to the observed slight long-term polarization drift. The polarization drift due to temperature stabilization could be reduced significantly using a temperature stabilized environment for the non-polarization-maintaining fiber free-space optics in the APC transmitter and receiver, respectively. Moreover, this could enable moving to an all-fiber measurement system using fiber polarizers and fiber squeezers in the APC receiver which would reduce loss and enable reduction of the APC reference lasers, thereby enabling reducing Raman noise.

During our classical testing of the APC, we investigated the signal wavelength dependence of the APC compared to the fixed references. For larger separations there was a noticeable decrease in polarization stabilization. This effect is also likely a contributing factor to the slight drift in the relative fidelity in the entanglement distribution and EAPT with 5-km fiber spool. We think there is a path toward improving the resilience of this APC system to PMD by the use of two sets of references which straddle the signal to be stabilized. By averaging measured polarization changes, we hypothesize that improved control can be achieved which have preliminary validation in some rough simulations but more analysis and testing is needed in further work. Nonetheless, in this work, we show PMD has only a minor affect when the reference and signal are separated by a few 100-GHz channels or less and when the fibers are on the few km-scale or somewhat longer.

{Finally, although this implementation and demonstration has been focused on fiber-based channels and devices, our methods (and their benefits) are not limited to fiber-based systems. In fact, there are small free-space subsystems in our transmitter and receiver systems. Moreover, the fiber-based or fiber-coupled critical components we used, namely, the acousto-optic modulator, heterodyne detector, variable-waveplate (fiber squeezer) have free-space versions, though in the case of the fiber squeezer it is a different physical mechanism producing the retardance (i.e., liquid crystal, electro-optic modulator, elasto-electric-optic modulator, etc.). Even the WDMs are micro-free-space optical systems with free-space filters inside. Just semi-minor changes would be required to apply these methods to free-space systems. But fiber-coupling into a fiber-based receiver is recommended primarily since polarization-maintaining fiber-based heterodyne detection has no issues with mode-matching and alignment that a free-space implementation would need to deal with by achieving and maintaining precision alignment. Furthermore, in free-space PMD is much less of a problem compared to in fiber leading to better performance. Finally, we would like to note that for links involving relative motion of reference frames (e.g., satellite to ground), Doppler shifts on the lasers would need to be accounted for and likely compensated for in the heterodyne detection.

Moreover, this compensation system is a closed-loop system which uses its own reference signals to produce corrections and does not rely on application-specific signals (i.e., quantum-bit error rate from a quantum key-distribution system); this enables our APC to encapsulate the channel and provide polarization channel stabilization which is fairly application agnostic.}

\appendix
\renewcommand{\thesection}{Appendix \Alph{section}}

\section{APC procedures}
\label{app:APCproc}
For calibration, here we assume the case of using a four fiber-squeezer device (alternating between H/V and D/A basis), a 50/50 splitter, polarization controllers, and polarizers. In this example, channel 1 fiber squeezer (Ch-1) is in the H/V basis, Ch-2 is in the D/A basis, Ch-3 is the H/V basis, and Ch-4 is the D/A basis. 

Since the single-mode fiber between the VWPs and the measurements will affect the polarization, calibration and stability of the polarization in that fiber is crucial. Note that polarization-maintaining fiber is not acceptable between the VWP and the measurements because it is not truly polarization-maintaining for all polarizations but only for certain input polarizations. Ideally, the VWPs and measurements could be all in fiber for robustness and compactness. In practice, we found the temperature sensitivity of such a system limits the calibration stability to the order of minutes without passive isolation and active temperature stabilization. Instead, we employed a free-space polarization measurement system as described above which afforded calibration over about 24 hours or more and additionally enabled precise mechanical control of the waveplate angles. It is also important the reference polarizations are stable as well. In practice, we found the wave plates to be much less temperature sensitive so all that was needed was to use a vented lid in the rack-mounted box housing the transmitter equipment.

Measurement calibration is achieved by taking advantage of the lack of modulation for polarizations which are in the basis of a VWP. The calibration procedure can be generalized from fiber squeezers to other arrangements of VWP; the calibration procedure is as follows:

\begin{enumerate}
    \item Adjust the polarization controller of the D-measurement so that back-propagated light from the D measurement (through the polarization-maintaining fiber-optic circulator, Thorlabs CIR1550PM-APC) is not polarization modulated by Ch-4. This combines three uses of the polarization controller: (a) make R state for back-propagation, (b) final QWP of compensating elements, and (c) undoing the fiber transformation so R state is along axis of squeezer.
    \item Using same back-propagated light, calibrate Ch-3 around the approx. $\pi/2$ or $-\pi/2$ phase shift to maximize modulation from Ch-2. This calibrates Ch-3 to be the QWP needed to rotate Ch-4 into the R/L basis.
    \item Adjust the polarization controller of the D-measurement so that back-propagated light from the D measurement is not polarization modulated by Ch-2.
    \item Adjust the polarization controller of the H-measurement so that back-propagated light from the H measurement is not polarization modulated by Ch-1.
\end{enumerate}

To redirect the receiver local oscillator for back-propagation during calibration, we used a patch panel as a switch board to achieve sufficient isolation. For the fiber squeezer polarization modulation signal, we used a 10~Hz 4~V$_{\text{pk-pk}}$, 2-V offset sine wave. To generate the signal on Ch-1 and Ch-2, the Moku:Go used for PID would be temporarily reprogrammed as a waveform generator. To generate the signal on Ch-4, the Moku:Go used for digital filtering of Ch-4 and PID for Ch-4 (and Ch-3 output), the digital filter would be temporarily reprogrammed as a waveform generator for Ch-4.  We measured the polarization modulation of the back-propagated light through a fiber polarizer with a power meter (Thorlabs). The peak-peak variation of the power-meter analog output was analyzed by an oscilloscope (Moku:Go available input since Ch-3 was constant with no PID). To reveal the maximum peak-peak variation present, a polarization controller, before the fiber polarizer, was adjusted as needed throughout the calibration steps. The peak-peak variation was minimized by an operator issuing computerized commands to rotate the relevant-measurement motorized wave-plates, iterating between the wave plates when a local minima was reached. This calibration could be completely automated if desired using an automated optical switch with sufficient isolation and a script reprogramming the Moku:Go, querying for the peak-peak variation, and adjusting the measurement wave-plates using the algorithms described.

This calibration procedure has been tested and verified to calibrate the path’s polarization transformation for control using the base algorithm described above without the slope sign included and we do not expect any changes for the slope-sign variation. The measurements were able to be calibrated within a few tenths of a degree using the motorized wave-plates. It is important that the calibration is stable for the duration of the polarization control. This fiber can be relatively short and coiled to a small space for physical securing from movement and temperature stabilization (if needed) which should result in long-term polarization stability. In the rack-box the receiver optical equipment resided in, we found it necessary to use foam insulation to passively stabilize the temperature of the fiber between the fiber squeezer and the free-space measurement system, including the OADMs; this resulted in stability within a few degrees of waveplate rotation over 24 hrs after a several-hour warm-up period. For longer-term operation, this calibration could be periodically checked and automatically adjusted using optical switches and/or circulators as well as automated fiber polarization controllers, e.g., fiber squeezers.

With the measurements calibrated, the LO could be redirected back to serve the heterodyne detection. Before closing the PID loops, the final task is to calibrate the signs, setpoints, and loop gains. The nature of the base PID algorithm is such that the feedback error signs of the different loops are not all the same. In practice, we had H/V-basis (Ch-1) and D/A-basis (Ch-2) loops have the same sign and R/L-basis (Ch-4) loop have different sign but simulation revealed multiple different sign combinations worked; the sign choice just affects the stabilized location on the sphere (e.g., R vs L). Obviously, this is taken care of automatically with the PID-with-slope-sign algorithm. To calibrate the setpoints, a polarization controller external to the receiver can be adjusted to find the maximum power received by each RF power detector (which is the minimum voltage due to its negative response slope). The algorithms dictate the ideal setpoint to be 50\% of the maximum signal power (e.g., the overlap of R onto H). The setpoints determine the location where the references are stabilized on the sphere and could be adjusted to stabilize the references to different places on the sphere and thereby stabilized the polarization channel to somewhat different transformations. Also, it should be considered that due to shot noise, a 3-dB reduction in the heterodyne signal peak will result in less than a 3-dB total power change; the difference between the two is dictated by the SNR to shot noise and filtering bandwidths. 

The loops gains were tuned to balance several factors. In general, the derivative was not used due to its propensity to cause instability and provide minor improvements for this sort of signal tracking. The integral term provided the primary control with the proportional term adding small faster corrections. To achieve minimal error while stabilizing the signal, the proportional and integral gains can be turned up until self-oscillation occurs then dialed back until somewhat lower than self-oscillation occurs. In our case, with the imperfect laser frequency stability, sometimes we tuned the PID to be slower and average over the power-variation of the frequency drift (e.g., Moku:Go PID proportional gain = 1 dB and integral corner frequency 1 Hz) which provided the most precise control we measured and was sufficient for in-ground fiber. For testing faster polarization drift artificially induced, we increased the integral corner frequency (up to 1~kHz in this work) but we did not find much control improvement by increasing the proportional gain though further testing and refinement may find more optimal tunings. Moreover, if there {i}s drift at {frequencies near} the edge of the PID bandwidth that can cause oscillation from 180$^{\circ}$ phase lag. To help dampen that oscillation (which did happen sometimes due to the frequency-drift-induced power variation when using 1-10 Hz integration bandwidth), we found it helpful to have the integral terms have slightly different bandwidth for each loop, e.g., 3~Hz, 2~Hz, and 1~Hz or 1~kHz, 800~Hz, and 600~Hz {to dampen oscillation}.

\section{{APC implementation details}}
\label{app:APCdetails}
{We will now describe experimental details of our implementation} for complete polarization channel stabilization which is depicted in Fig.~\ref{fig:apcsetup}. Starting at the transmitter, we prepare two reference signals by filtering out the laser pedestal of a power- and frequency-stabilized laser (Pure Photonics PPCL551, PPCL590) and splitting the light along two paths. One path is attenuated {(to balance power with the other path)} and right-circularly polarized (as measured after the 50/50 beamsplitter) while the other path is frequency-shifted by $X=200$~MHz with an acousto-optic modulator (Aerodiode 1550AOM-2)  before being and horizontally polarized. The two paths are mixed on a 50/50 beamsplitter and half the light is collected into single-mode fiber. The collected light is power-stabilized (Thorlabs EVOA1550A) and attenuated (Thorlabs VOA50-APC). Using a WDM, the reference signals combined with the (quantum) signal to be stabilized. 

After going through the channel, the quantum signal and reference signals propagate through the polarization compensation device, in this case, piezo fiber squeezers (Luna Innovations/General Photonics PCD-M02-4X-7-FC/APC) chosen for variable-waveplate nature, relatively high-bandwidth, and low fiber-connection insertion loss. Using 100-GHz optical add-drop multiplexers (OADM, AC Photonics), the reference signals are demultiplexed from the quantum signal. An optional 100-GHz DWDM can also be used to provide additional filter if needed, which we used. This OADM filtering makes a total signal insertion loss of 1.5-2 dB (2-2.5 dB with an additional DWDM). 

 The reference signal is split using a 50/50 beamsplitter and each path measures a different polarization (horizontal and diagonal) using a hybrid Fiberbench/cage system (Thorlabs) housing zero-order half- and quarter-wave plates mounted in motorized rotation mounts (Thorlabs K10CR2). The polarization reference light exiting each polarization measurement are each directed toward a different heterodyne spectrometer {for detection}. To achieve heterodyne detection, another identical frequency-stabilized laser with $+6$-dBm power is split and used as the local oscillator for each heterodyne spectrometer using polarization-maintaining balanced detection (Thorlabs PN1550R5A2 and PDB835C-AC). 

The frequency-stabilized lasers used were well calibrated by the manufacturer so that when set to the same wavelength they would be within a few hundred MHz which is within the balanced detector's electrical bandwidth. To do further fine-tune adjustments, part of the heterodyne signal from Detector 1 was analyzed in the frequency domain by an oscilloscope (Keysight/Agilent MSOX4104A). The frequency-domain analysis does not distinguish between positive and negative frequencies but when both received signal peaks move positively to higher frequencies when reducing the local oscillator frequency indicates that the received references are higher in frequency than the local oscillator as designed. This frequency-stabilized laser at the receiver is then fine tuned so the peak of the unshifted reference has an optical frequency offset of approximately $Y=70$~MHz that matches a pre-determined fraction of the balanced detector’s electrical bandwidth $B$, discussed below.  The electrical output of heterodyne Detector 1 (which detects the vertically polarized light) is bandpass filtered to transmit the electrical signals near frequency $Y$. Whereas, heterodyne Detector 2 (which detects the diagonally polarized light) is bandpass filtered to transmit the electrical signals {near frequency $Y$, and separately,} near frequency $Y + X$. 

There are several constraints to satisfy for this system to work as desired. First, $Y + X$ should be roughly equal to or less than $B$ so the desired signals are within the balanced detector’s bandwidth(s). In general, the two heterodyne detectors do not need to use detectors with the same bandwidth and this could be advantageous since the detector bandwidth effectively acts like a low-pass filter. Also, the lasers have laser frequency drift $\nu$ and linewidth $l$. Heterodyne detection involves the convolution of two lasers so we expect a total drift of approximately $(\nu \sqrt{2})(l\sqrt{2})=2\nu l$. Want $X> 2\nu l$ so the signals can be easily separated with fixed filters. For example, in our demonstration, we choose $\nu<125$~MHz, $l=10$~kHz, $B=500$~MHz, $Y=70$~MHz, $X=200$~MHz, Detector 1 and Detector 2a band-pass filter {($F_{\text{B1}}$ and $F_{\text{B2a}}$)} passband = 10-200~MHz (Mini-circuits SLP-200+ and ZKL-1R5+), {Detector 1 and Detector 2a low-pass filter ($F_{\text{L1}}$ and $F_{\text{L2a}}$) passband = 0-200~MHz (Mini-circuits SLP-200+), Detector 2b high-pass filter ($F_{\text{H2b}}$) passband > 200~MHz (Mini-circuits SHP-250+),} and Detector 2b band-pass filter {($F_{\text{B2b}}$)} passband = 200-400~MHz (Mini-circuits SHP-250+ and SLP-550+). 

In practice, we found that the frequency drift combined with frequency-dependent RF gain of the signal chain resulted in intolerable RF power variation of the measured signals. To reduce this variation to tolerable levels, we employed a soft frequency stabilization of the relative frequency of the LO and reference lasers. To measure the relative frequency, we split some of the power from Detector 1 (Mini-circuits Z99SC-62-S+), applied a slow-pass filter to isolate the lower-frequency peak (2x Mini-circuits SLP-200+) and amplified the signal (Mini-circuits ZKL-1R5+) before counting zero-Voltage threshold crossings to approximately determine the frequency (Keysight/Agilent MSOX4104A). This measurement had reduced signal-to-noise ratio (SNR) due to the heterodyne shot noise {so we increased} the received reference power to be high enough (-50 dBm) for sufficient SNR on the frequency measurement. At this power level, the heterodyne signal had an SNR of 20-30 dB above shot noise so the reference power could be reduced with a better frequency measurement. The frequency measurement is queried via a network connection to a computer (Raspberry Pi) running a python script to stabilize the measured frequency with a setpoint using PID which gives a correction to the LO via serial command to adjust the fine-tune laser frequency adjustment (thermally actuated with bandwidth of about 10 Hz, Pure Photonics). Furthermore, we optimized the PID setpoint ($Y$) so that $X+Y$ had the minimum frequency-dependent RF gain within the constraint of stable PID performance considering the noisy frequency measurement. 

Without this frequency stabilization, the heterodyne signal had about 1-Hz frequency drift of about 200~MHz wide and $<< 1$-Hz frequency drift of about 200 MHz as well due to the independent laser drifts. Together, the laser was seen to drift, even momentarily, out of the 0-200 MHz desired range. With the frequency stabilization though, the heterodyne signal was stably maintained at the 70-MHz setpoint with about 5-10 MHz residual standard deviation for more than 24 hours. If desired, the residual variation could be further reduced by incorporating a faster bandwidth frequency adjustment (e.g., the 100-kHz piezo-based frequency-modulation analog input of the laser) which should be sufficient for the about 10-Hz fastest drifts seen sometimes. 

The filtered heterodyne detection outputs are also amplified (Mini-circuits ZKL-1R5+) then filtered again (to improve signal-to-noise ratio). The output of which has its RF power linearly detected (Mini-circuits ZX47-60-S). The output of the power detector is low-pass filtered (10-kHz 8-th order Butterworth digital low-pass filter, Liquid Instruments Moku:Go); this filtering reduces noise from the power detector but this also purposefully limits the feedback bandwidth of the polarization stabilization. The detected power signals can be used by a feedback algorithm which actuates a polarization compensation device (i.e., fiber squeezers module) to stabilized the polarization. 

\section{Additional classical test signal data}
\label{app:addclasstest}
\begin{figure}
\centerline{\includegraphics[width=1\textwidth]{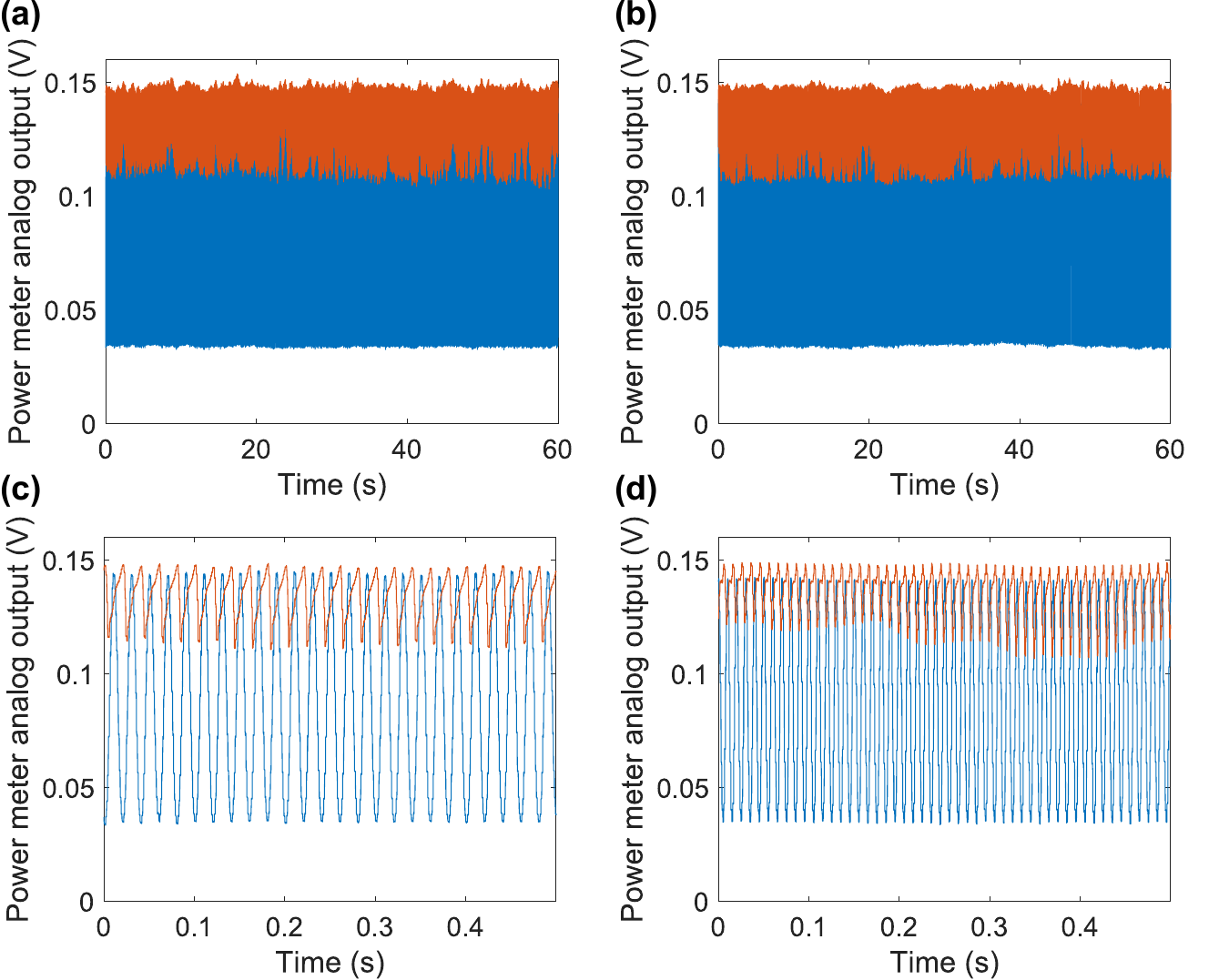}}
\caption{C-band Classical test signals with(out) C-band APC. APC reference laser wavelength was 1558.2 nm. Test signal wavelength was 1559.75 nm, about 2 100-GHz channels away.   Test signals after 50-Hz (a) and 100-Hz (b) induced polarization drift without APC (blue) and with APC (orange) using about 1 kHz integration bandwidth. (c) and (d) are zoomed in versions of (a) and (b), respectively.}
\label{fig:CbFSFGteststoofast}
\end{figure}

To further test the APC bandwidth for a 1-kHz PID integral bandwidth, we also applied 50 Hz and 100 Hz induced polarization variation in Fig.\ref{fig:CbFSFGteststoofast}. Since the speed of variation is a greater fraction of the PID integral bandwidth there is less correction applied by the APC but still there is some correction.
\section{Partial entanglement-assisted process tomography comparison}
\label{app:partEAPTcomp}
\begin{figure}
\centerline{\includegraphics[width=1\textwidth]{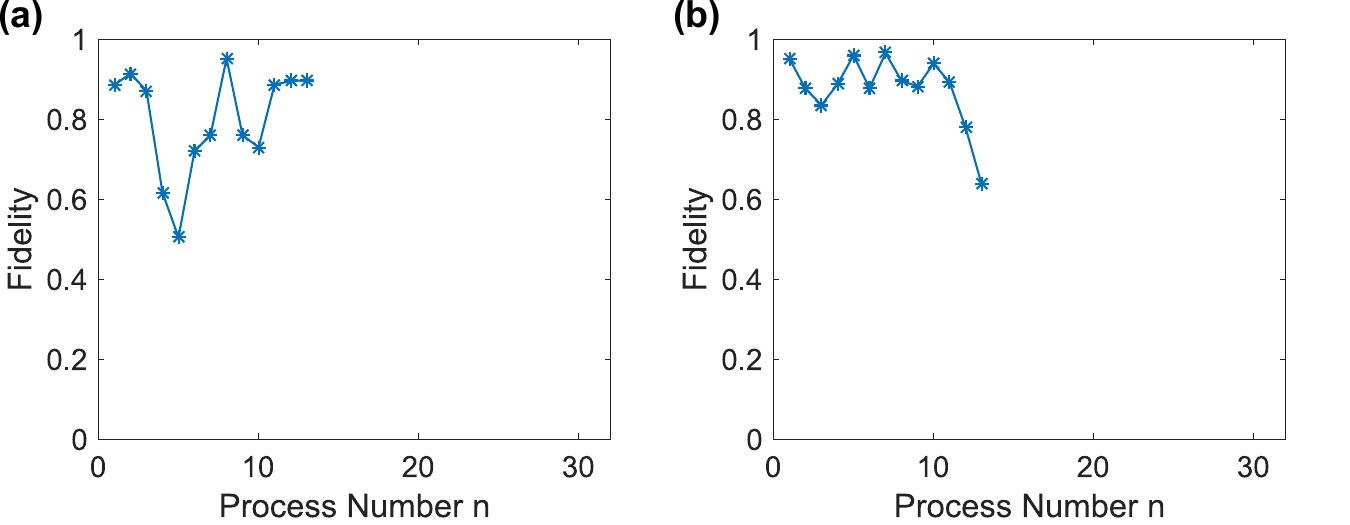}}
\caption{Relative process fidelities (RPF) for two successive series of process tomographies. (a) RPF between $\mathfrak{C}_n$ and $\mathfrak{C}_{n+1}$ and (b) RPF between  $\mathfrak{C}_1$ and $\mathfrak{C}_n$ without the APC enabled for both sides going through a different 5-km spool. Integration time per measurement setting was 5~s using a 660-ps coincidence window, collecting an average  $4.6x10^4$  coincidence counts per 36-setting tomography.}
\label{fig:EAPTpartcomp}
\end{figure}

In this configuration, we measured a series of tomographies without the APC enabled where both photons went through a different 5-km spool. The RPT between $\mathfrak{C}_n$ and $\mathfrak{C}_{n+1}$ are shown in Fig.~\ref{fig:EAPTpartcomp}(a) and between $\mathfrak{C}_1$ and $\mathfrak{C}_n$ are shown in Fig.~\ref{fig:EAPTpartcomp}(b). The variation in RPT in Fig.~\ref{fig:EAPTpartcomp}(a) shows there is significant polarization drift during a single measured tomography (about 10 minutes time). Whereas, the variability of RPT in Fig.~\ref{fig:EAPTpartcomp}(b) shows polarization drift between the measured tomographies as well. In this case, we have an upper-bound fidelity variation for Fig.~\ref{fig:EAPTresults}.

\section{PID output logging during entanglement distribution}
\label{app:PIDoutentdist}
\begin{figure}
\centerline{\includegraphics[width=1\textwidth]{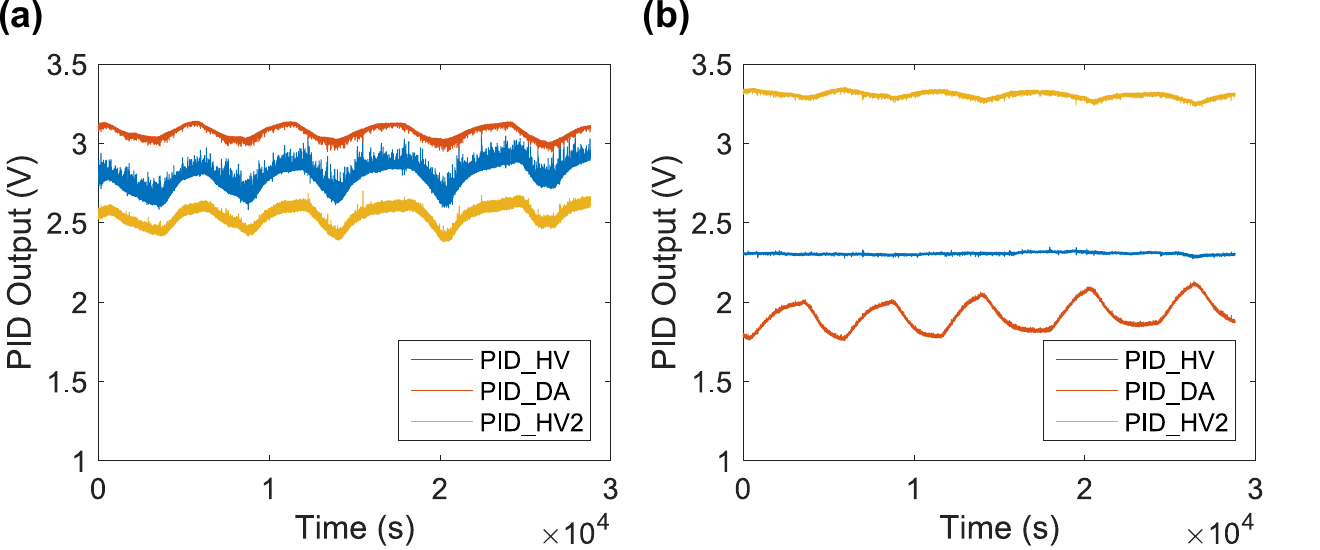}}
\caption{APC PID output logging during entanglement distribution. The PID outputs shown here were continuously logged for the last 8 hrs of the entanglement distribution testing with the end matching up with the end of Fig.~\ref{fig:EntDistdata}. (a) Outputs from the C-band APC traversing from UTC to the Qubitekk quantum node. (b) Outputs from the L-band APC traversing from UTC to the Broad St. quantum node.}
\label{fig:entdistPIDlog}
\end{figure}

The PID outputs were logged during several parts of the entanglement distribution. The longest continuous span is shown in Fig.~\ref{fig:entdistPIDlog}. It is noticeable that the C-band APC outputs are noisier than the L-band APC outputs. We attribute this to increased frequency drift of the C-band reference/LO lasers compared to the L-band reference/LO lasers.

\section{Polarization characterization details}
\label{app:polchar}
To characterize the bandwidth of polarization variation, we transmitted a 1550-nm polarized laser (Pure Photonics PPCL550) on the EPB quantum network from the Qubitekk quantum node to a node at the Broad St. quantum hub via the node at the University of Chattanooga to pass through both links planned for the entanglement distribution test discussed later. The received polarization was measured using a fiber polarization beamsplitter (Thorlabs PBC1550SM-FC) and power meters (Thorlabs PM101A, S154C). 

The analog power meter outputs were sampled at 100 MHz by an analog-to-digital converter (Analog Devices AD9254) and interfaced to a system on chip (field-programmable gate array, FPGA, with co-located hard processor, Intel Cyclone V SoC) by a demonstration board (Terasic ADCSoC). In the FPGA, sets of $2^8$ 100-MHz samples are continuously accumulated to produce 1.56-MHz samples. Similarly, $2^{16}$ 100-MHz samples are continuously accumulated to produced 1.56-kHz samples. To limit saved data non-volatile memory requirements, every $2^{32}$ 100-MHz clock cycles (about 43 s), $2^{10}$ sets of $2^8$ 100-MHz samples and $2^{12}$ 1.56-MHz samples are placed in first-in-first-out (FIFO) buffers, respectively. Similarly, 1.56-kHz samples are continuously placed in a FIFO buffer. Arbitration is used to combine FIFO buffers into a single FIFO buffer whose data is transferred to a buffer located in the volatile random-access-memory (RAM) accessible by the linux program running on the hard processor. 

A threaded linux program can check a counter in the FPGA to ensure program access to different parts of the RAM buffer at the right times. One thread reads a new set of 100-MHz and 1.56-MHz samples from the RAM buffer every 43~s. After converting the accumulated samples to average samples by division, this thread normalizes the samples by dividing one of the sampled inputs by the sum of both inputs after shifting the samples by half the range of the ADC ($2^{13}$) to make the normalized samples always positive then FFTs are calculated on each set of samples. The average and standard deviation of $2^{10}$ FFTs of each sample rate are calculated then saved to non-volatile memory. 

The other thread reads a new set of $2^{12}$ 1.56-kHz samples about every 2.68~s (because $2^{12}*2^{16}*1/(100\text{ MHz}) = 2.684$ s). The new samples are converted from accumulated to averaged after division by $2^{16}$ and then normalized. Every $2^{10}$ 1.56-kHz samples are averaged together to create a 1.56-Hz sample; similarly, every $2^{10}$ 1.56-Hz samples are averaged together to create a 1.56-mHz sample (which is immediately appended to a file in non-volatile memory). After $2^{12}$ new 1.56-kHz samples are processed, the thread then calculates the fast Fourier transforms (FFT) of each set. The average and standard deviation of $2^{5}$ 1.56-kHz FFTs are calculated then saved to non-volatile memory. After $2^{12}$ new 1.56-Hz samples are collected, the FFT is calculated and saved to non-volatile memory.

\begin{backmatter}
\bmsection{Funding}
Content will be automatically generated by Publisher.

\bmsection{Acknowledgments}
J.C.C. led the project and demonstration design, as well as devised, constructed, and tested the APC, and also led the results{,} analysis{,} and paper composition. M.A. developed the entanglement source, tomography systems, and upgraded time-tagger hardware, as well as helping with results{,} analysis{,} and paper composition. K.R. provided experimental assistance during the Chattanooga testing and demonstration. T.L. facilitated laboratory usage for the Chattanooga testing and demonstration and assisted with paper composition. M. K. provided helpful managerial oversight and assisted with paper composition.
Most of this work was performed at Oak Ridge National Laboratory, operated by UT-Battelle for the U.S. Department of Energy under contract no. DE-AC05-00OR22725. J.C. and M. A. acknowledge research sponsored by the Laboratory Directed Research and Development Program of Oak Ridge National Laboratory, managed by UT-Battelle, LLC, for the U. S. Department of Energy. J.C. acknowledge the polarization characterization hardware development sponsored by 
U.S. Department of Energy, Office of Science, Advanced Scientific Computing Research, under the Performance integrated quantum scalable internet program (Field Work Proposal ERKJ432). T.L. and K.R. acknowledge the support of the UTC Quantum Initiative in contributing to this work.
\bmsection{Disclosures}
The authors declare no conflicts of interest.

\bmsection{Data Availability Statement}
Data underlying the results presented in this paper are not publicly available at this time but may be obtained from the authors upon reasonable request.

\end{backmatter}



\end{document}